\documentclass[twocolumn]{aastex63}
\usepackage{lineno}
\usepackage{tabularx}
\usepackage{scalerel}
\graphicspath{{./}{figures/}}

\begin{document}
\title{Revealing AGNs Through TESS Variability}

\author[0000-0003-0660-9776]{Helena P. Treiber} 
\affiliation{Institute for Astronomy, University of Hawai`i, 2680 Woodlawn Dr., Honolulu, HI 96822, USA}
\affiliation{Department of Physics and Astronomy, Amherst College, C025 New Science Center, 25 East Dr., Amherst, MA 01002-5000, USA}
\author[0000-0001-9668-2920]{Jason T. Hinkle} 
\affiliation{Institute for Astronomy, University of Hawai`i, 2680 Woodlawn Dr., Honolulu, HI 96822, USA}
\author{Michael M. Fausnaugh}
\affiliation{Department of Physics and Kavli Institute for Astrophysics and Space Research, Massachusetts Institute of Technology, Cambridge, MA 02139, USA}
\author[0000-0003-4631-1149]{Benjamin J. Shappee} 
\affiliation{Institute for Astronomy, University of Hawai`i, 2680 Woodlawn Dr., Honolulu, HI 96822, USA}
\author[0000-0001-6017-2961]{Christopher S. Kochanek} 
\affiliation{Department of Astronomy, The Ohio State University, 140 West 18th Avenue, Columbus, OH 43210, USA}
\author[0000-0001-5661-7155]{Patrick J. Vallely} 
\affiliation{Department of Astronomy, The Ohio State University, 140 West 18th Avenue, Columbus, OH 43210, USA}
\author[0000-0002-4449-9152]{Katie Auchettl}
\affiliation{School of Physics, University of Melbourne, Victoria 3010, Australia}
\affiliation{Center for Cosmology and Astroparticle Physics, The Ohio State University, 191 W.~Woodruff Avenue, Columbus, OH 43210, USA}
\author[0000-0001-9206-3460]{Thomas W.-S. Holoien}
\altaffiliation{NHFP Einstein Fellow}
\affiliation{The Observatories of the Carnegie Institution for Science, 813 Santa Barbara Street, Pasadena, CA 91101, USA}
\author[0000-0003-3490-3243]{Anna V. Payne}
\altaffiliation{NASA Graduate Fellow}
\affiliation{Institute for Astronomy, University of Hawai`i, 2680 Woodlawn Dr., Honolulu, HI 96822, USA}
\author[0000-0001-9203-2808]{Xinyu Dai}
\affiliation{Homer L. Dodge Department of Physics and Astronomy, University of Oklahoma, Norman, OK 73019, USA}

\begin{abstract}
    We used Transiting Exoplanet Survey Satellite (TESS) data to identify 29 candidate active galactic nuclei (AGNs) through their optical variability. The high-cadence, high-precision TESS light curves present a unique opportunity for the identification of AGNs, including those not selected through other methods. Of the candidates, we found that 18 have either previously been identified as AGNs in the literature or could have been selected based on emission-line diagnostics, mid-IR colors, or X-ray luminosity.  
    AGNs in low-mass galaxies offer a window into supermassive black hole (SMBH) and galaxy co-evolution and 8 of the 29 candidates have estimated black hole masses $\mathrm{\lesssim 10^{6} M_{\odot}}$.
    The low-mass galaxies NGC 4395 and NGC 4449 are two of our five ``high-confidence" candidates.
    By applying our methodology to the entire TESS main and extended mission datasets, we expect to identify $\sim$45 more AGN candidates, of which $\sim$26 will be new and $\sim$8 will be in low-mass galaxies.
    
\end{abstract}


\section{Introduction} 
\label{s.intro}
Most massive galaxies contain supermassive black holes \citep[SMBHs,][]{Kormendy1995, Richstone1998}.
Active galactic nuclei (AGNs), where the SMBH is accreting, are seen in 1-5\% of galaxies in the local universe \citep[e.g.,][]{Ho2008,haggard10, lacerda20,Mishra2020,Yuk2022}. AGNs inform our understanding of galaxy and SMBH co-evolution and provide a detailed look at the physics of accretion \citep[see][for a review]{Heckman2014}. 

AGNs are commonly selected using emission-line diagnostics \citep[e.g.,][see Section \ref{s.emission}]{Baldwin1981, Veilleux1987, Kewley2006, Fernandes2011}, color \citep[e.g.,][see Section \ref{s.wise}]{Koo1988,Fan1999,Richards2001,Lacy2004,Stern2012,Assef2013}, radio \citep[e.g.,][]{Baade1954,Tadhunter2016}, and X-ray luminosity \citep[e.g.,][see Section \ref{s.xray}]{Elvis1978,Mendez2013}.
These methods are sensitive to different AGN populations \citep[e.g.,][]{Hickox2009}. For example, although spectroscopic methods can be used in the identification of AGNs in low-mass galaxies, the low metallicity and higher star formation rates in these galaxies can complicate line ratio diagnostics \citep[e.g.,][]{Trump2015}.

A key feature of AGNs is their stochastic variability across the electromagnetic spectrum. 
As a result, variability selection is one of the most promising ways of identifying actively accreting BHs.
In particular, emission-line diagnostics miss the majority of variability-selected AGNs in low-mass galaxies \citep[][]{Baldassare2020,BurkeDES,Burke2021,Ward2021,Latimer2021,Yuk2022}.
The fraction of low-mass galaxies with an AGN (i.e., the active fraction) is not well-constrained, but has been predicted to be lower than at higher masses \citep[e.g.,][]{Pacucci2021}. However, \cite{BurkeDES} found a consistent variable active fraction across SMBH mass. In addition to enlarging the known population of low-mass AGNs, the light curves from variability selection can be used to probe the physics of accretion. In particular, photometric reverberation mapping can be used to measure accretion disk sizes \citep[e.g.,][]{Haas2011,Shappee2014, fausnaugh18}.

AGN variability is often described by a damped random walk \citep[DRW; e.g.,][]{Kelly2009, Kozlowski2010, MacLeod2010}. \cite{MacLeod2010}, \cite{Kelly2009}, and \cite{Burke2020} all find that the characteristic timescale of the DRW model ($\tau_{DRW}$) becomes shorter for lower-mass SMBHs. 
An AGN is unlikely to vary faster than the light-crossing time of its accretion disk, which is larger for a higher-mass black hole. 
Thus, regardless of the efficacy of the DRW model, we expect lower-mass SMBHs to vary more rapidly \citep[e.g.,][]{Xie2005}. 

Thanks to the
high cadence of the Transiting Exoplanet Survey Satellite (TESS; \citealp{Ricker2014}), we can now search for optical variability on $\sim$hour timescales.
\cite{Burke2020} showed that the well-known Seyfert 1 in the dwarf galaxy NGC 4395 \citep[][]{Filippenko1989} optically varies with a $\mathrm{\tau_{DRW}}$$\sim$2.5 days. 
This result suggests that the cadence, precision, and all-sky coverage of TESS should make it a useful tool in the study of variability from AGNs in low-mass galaxies. Furthermore, the $\tau_{DRW}$ of NGC 4395 indicates that the $\mathrm{\tau_{DRW}-M_{BH}}$ relation extends several orders of magnitude down to this $\sim$$10^5 \mathrm{M_{\odot}}$ BH. 
\cite{Burke2021} corroborated this result by adding several AGNs with variability observed by other optical telescopes to the relation. 

In this paper, we search for AGNs using TESS light curves. 
During its primary two-year mission, the main goal of TESS was to identify transiting exoplanets around M dwarf stars \citep[e.g.,][]{gilbert20}. 
The resulting photometric precision and high cadence make TESS an excellent tool for a much broader range of time-domain studies \citep[e.g.,][]{Holoien2019,Mishra2020,Payne2021,Fausnaugh2021,Vallely2021, zeldes21, Hinkle2022a, hinkle22, payne22, way22}.
Using four $24^{\circ} \times 24^{\circ} $ FOV cameras, TESS repeatedly images one of 26 sectors (13 per hemisphere) for $\sim$27 days at a time. The wide 600-1000 nm bandpass is centered on the Cousins $I$ band, with an effective wavelength of 745.3 nm \citep{Ricker2014,Rodrigo2020}. 
We focus on the mission's first two years, when TESS took full-frame images (FFIs) every 30 minutes. 

\begin{figure*}
    \centering
    \includegraphics[width=2.0\columnwidth]{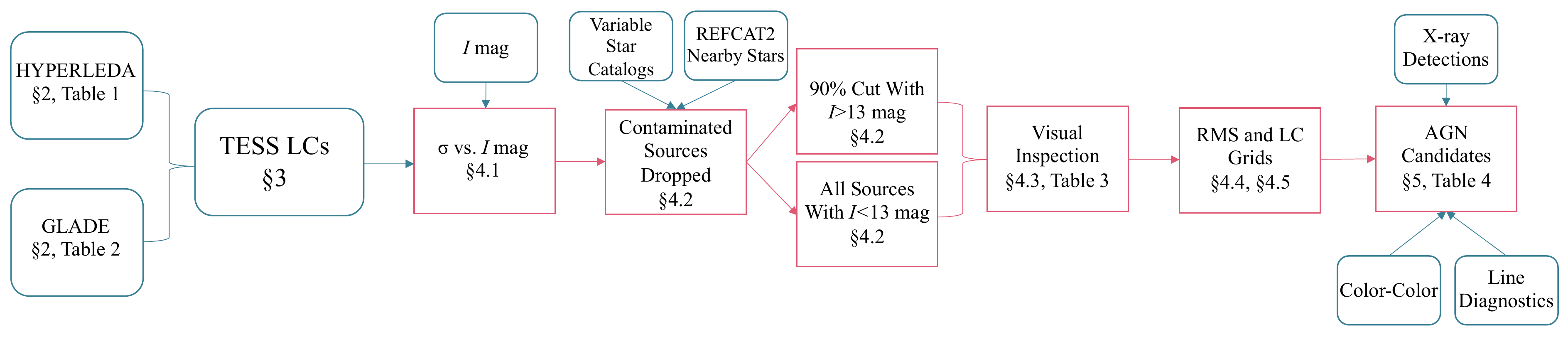}
    \caption{Flow chart outlining steps from sample selection to AGN candidate identification. Steps that introduce data are shown with rounded blue outlines, while cuts are described in pink rectangles.}
    \label{fig:flow}
\end{figure*}

In Figure \ref{fig:flow}, we outline the steps of our search for AGNs using TESS.
In Section \ref{s.sample}, we explain our sample selection. We summarize the steps involved in TESS light curve generation in Section \ref{s.TESS}. We detail our methodology to account for stellar contamination and systematics, such as scattered light and spacecraft jitter \citep[e.g.,][]{Vanderspek2018}. Following variability cuts, we present the 29 AGN candidates in Section \ref{s.candidates}. 
In Section \ref{s.diagnostics}, we investigate archival data for the 29 AGN candidates. We demonstrate that up to 38\% of the AGN candidates can only be selected using variability.
We assume the same cosmological parameters as HyperLEDA \citep[][]{HyperLEDA} to convert between redshift and distance:
$H_{0} = 70 \: \mathrm{km \: s^{-1} \: Mpc^{-1}}$, 
$\Omega_{M} = 0.27$, and
$\Omega_{\Lambda} = 0.73$.
\section{The Sample}
\label{s.sample}
Our full sample comes from two sources: one from HyperLEDA \citep{HyperLEDA}, limited by TESS sector but not mass, and one from the Galaxy List for the Advanced Detector Era \citep[GLADE,][]{GLADE}, limited by mass but not sector.
We only included galaxies with $I < 16$ mag because statistical noise dominates the light curves of fainter galaxies.

The majority of our sample consists of galaxies cataloged in HyperLEDA.\footnote{http://leda.univ-lyon1.fr/} By compiling and processing several galaxy catalogs, HyperLEDA maintains a database of $\sim$5 million objects. 
The TESS light curves of the 137,695 HyperLEDA sources in our sample had already been generated, along with galaxies as faint as $I=20$ mag, using the techniques of \citet{Fausnaugh2021}.
The sample consists of sources in nine sectors (2, 3, 4, 5, 6, 7, 19, 20, and 21). Based on the light curves for the first few sectors, it became clear that scattered light presents a significant challenge, particularly for the fainter sources (i.e., $I > 16$ mag), for about half of each year \citep{Vanderspek2018}. 
Therefore, we only consider the light curves for sectors where this contribution is much lower: 5, 6, 7, 19, 20, and 21.
Table \ref{t.hyp} includes TESS-related information and relevant galaxy properties cataloged in HyperLEDA. 

For the GLADE sample, we considered galaxies likely to have lower-mass central BHs ($\mathrm{M_{BH}}$ $\mathrm{\lesssim 10^{7} M_{\odot}}$) based on their absolute $K$ magnitudes.
GLADE tries to achieve completeness by combining data from many other catalogs.  
In addition to the mass cut, we limited the redshift to $0.0001<z<0.5$ to minimize contamination by stars and higher-redshift quasars. 
GLADE does not report $I$ band magnitudes. Using GLADE galaxies also in the HyperLEDA catalog, we found that
\begin{equation}
    I = J \times 0.91 + 2.0
\end{equation}
was a reasonable proxy to use for our $I\lesssim16$ selection limit.
Although 5676 galaxies in GLADE satisfied our selection parameters, 1310 of them either fell between pointings in the TESS primary mission or on the edge of the detector.
Table \ref{t.sg} includes the 4366 low-mass sources from GLADE for which we could generate TESS primary mission light curves.

We estimated black hole masses for both the HyperLEDA and GLADE samples using the relation from \cite{Graham2007}, which has a scatter of $\sim$0.33 dex. For the GLADE sample, we cut at $\mathrm{M_{BH}}$ of $\mathrm{10^{7}M_{\odot}}$ to restrict the sample to low-mass black holes.
This empirical correlation used galaxies with $K$-band measurements for the spheroidal component of the galaxy and independently-estimated black hole masses. However, at this stage, we used the \cite{Graham2007} relation on the total galaxy absolute $K$-band magnitudes for two reasons.
First, the majority of galaxies in the GLADE sample have only photometric redshifts, making the distances and resulting BH mass estimates significantly uncertain ($\gtrsim$1 dex), especially at low redshift  \citep[][]{Bilicki2014,GLADE}.
Second, we lack morphological information for all of the GLADE and most of the HyperLEDA galaxies.
We treat the BH mass estimates using the \cite{Graham2007} relation as approximate upper limits.

\begin{figure*}
    \centering
    \includegraphics[width=2\columnwidth]{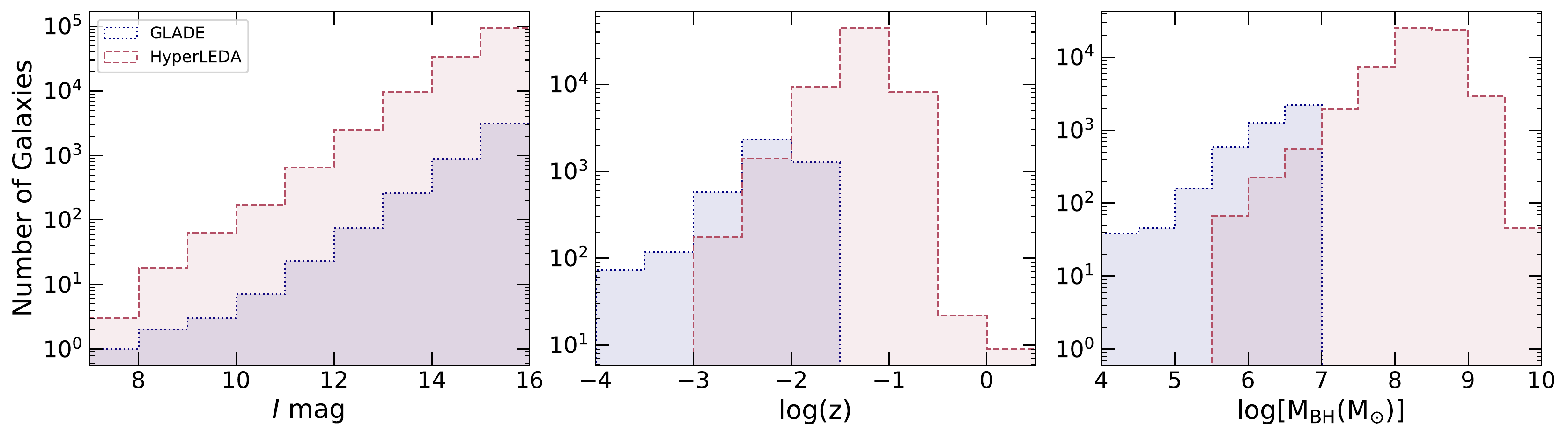}
    \caption{Histograms of the apparent $I$-band magnitude, redshift, and estimated black hole mass for galaxies in the HyperLEDA and GLADE samples. GLADE galaxies are shown by a navy dotted line and HyperLEDA uses a pink dashed line. 
    There are a few mass estimates below $\mathrm{10^{4} M_{BH}}$ but we doubt they are realistic.}
    \label{fig:hists}
\end{figure*}

For the final AGN candidates, we obtained morphological classifications and either redshift-independent distances or spectroscopic redshifts to re-estimate the candidate BH masses (see Section \ref{s.finalcands}). For four of the five candidates from the GLADE sample, the new estimate yielded lower BH mass estimates by over an order of magnitude. The major driver of this change was the incorporation of morphologies. The distributions of the samples in $I$ mag, redshift, and estimated black hole mass are shown in Figure \ref{fig:hists}.

\begin{deluxetable*}{lDDcccc}
\tablewidth{0.7\textwidth}
\tabletypesize{\footnotesize}
\tablecaption{HyperLEDA sample}
\tablehead{
\colhead{Name} &
\multicolumn2c{RA} &
\multicolumn2c{Dec} &
\colhead{\# of Sectors} &
\colhead{$I$} &
\colhead{$\mathrm{K_{abs}}$} &
\colhead{log($\mathrm{M_{BH}/M_{\odot}}$)} \\
\hline
\colhead{} &
\multicolumn2c{deg} &
\multicolumn2c{deg} &
\colhead{} &
\colhead{mag} &
\colhead{mag} &
\colhead{}
}
\decimals
\startdata
PGC2225524&122.867 &43.638 &1&15.9&$-$25.9&9.0\\ 
PGC871065&51.110 &$-$18.217 &1&15.2&\dots&\dots\\
PGC368549&113.691 &$-$59.807 &4&13.5&$-$25.0&8.7\\ 
PGC871073&35.172 &$-$18.217 &1&15.9&\dots&\dots\\
PGC871961&39.379 &$-$18.157 &1&15.5&$-$25.3&8.8\\
PGC558384&10.156 &$-$42.764 &1&16.0&\dots&\dots\\
PGC558887&24.735 &$-$42.733 &2&15.5&$-$23.8&8.2\\
PGC2228747&124.636 &43.771 &1&15.6&$-$26.6&9.2\\ 
PGC282848&108.633 &$-$68.873 &3&15.3&\dots&\dots\\
PGC638517&92.479 &$-$36.294 &1&15.9&\dots&\dots\\
\enddata 
\tablecomments{Identifying information and key values for the 137,695 galaxies in the HyperLEDA sample. Only a section of the table is shown here to demonstrate table formatting; the rest can be found in the ancillary files.}
\label{t.hyp} 
\end{deluxetable*}

\begin{deluxetable*}{DDccccc}
\tablewidth{0.7\textwidth}
\tabletypesize{\footnotesize}
\tablecaption{GLADE sample}
\tablehead{
\multicolumn2c{RA} &
\multicolumn2c{Dec} &
\colhead{\# of Sectors} &
\colhead{$I$} &
\colhead{$\mathrm{K_{abs}}$} &
\colhead{log($\mathrm{M_{BH}/M_{\odot}}$)}\\
\hline
\multicolumn2c{deg} &
\multicolumn2c{deg} &
\colhead{} &
\colhead{mag} &
\colhead{mag} &
\colhead{}
}
\decimals
\startdata
182.494 &46.457 &1&11.7&$-$19.2&6.5\\
30.321 &28.837 &1&12.9&$-$17.3&5.8\\
187.062 &11.790 &1&13.2&$-$19.6&6.7\\
190.210 &4.526 &1&16.0&$-$15.7&5.2\\
52.033 &51.938 &1&13.8&$-$13.7&4.5\\
44.361 &56.869 &1&15.3&$-$12.0&3.9\\
17.091 &59.856 &2&15.6&$-$9.8&3.1\\
4.766 &59.825 &3&15.5&$-$12.3&3.9\\
33.659 &64.210 &2&15.7&$-$11.6&3.7\\
345.433 &71.763 &2&15.4&$-$13.8&4.5\\
\enddata 
\tablecomments{TESS information and galaxy properties for the 4366 galaxies in the GLADE sample. Only a section of the table is shown here; the rest can be found in the ancillary files.}
\label{t.sg} 
\end{deluxetable*}
\begin{figure}
    \centering
    \includegraphics[width=\columnwidth]{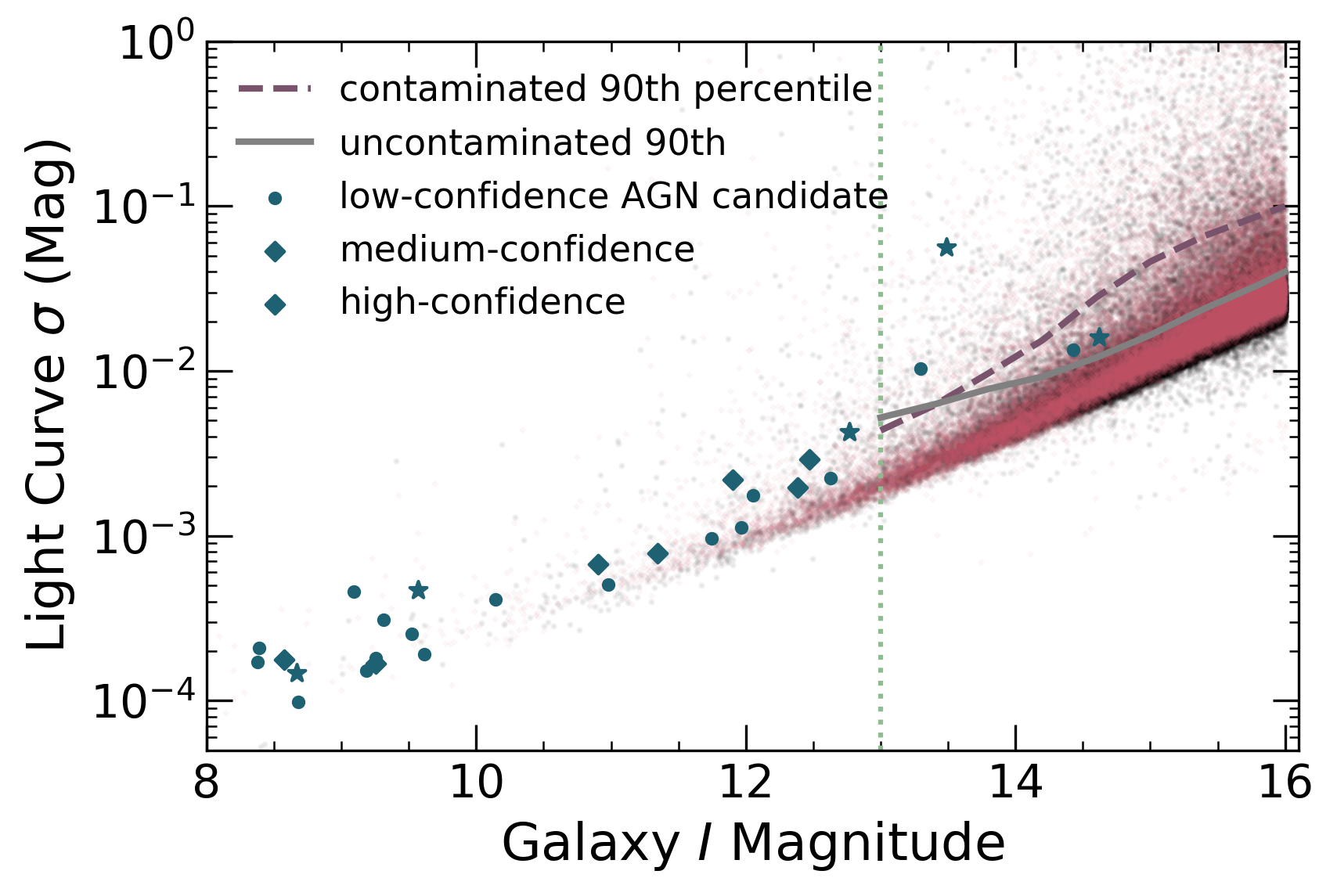}
    \caption{Standard deviation vs. magnitude for the whole sample, illustrating the increasing noise at fainter magnitudes. Sources found to be contaminated prior to visual inspection are shown in pink, and the rest are in black. The purple dashed line follows the $\mathrm{90^{th}}$ percentile for the contaminated sources only, while the grey solid curve follows the uncontaminated $\mathrm{90^{th}}$ percentile. Below $\mathrm{13^{th}}$ mag (shown by the vertical dotted line), we inspect all uncontaminated sources. Final AGN candidates are marked.}
    \label{fig:prec}
\end{figure}
\begin{figure}
    \centering
    \includegraphics[width=\columnwidth]{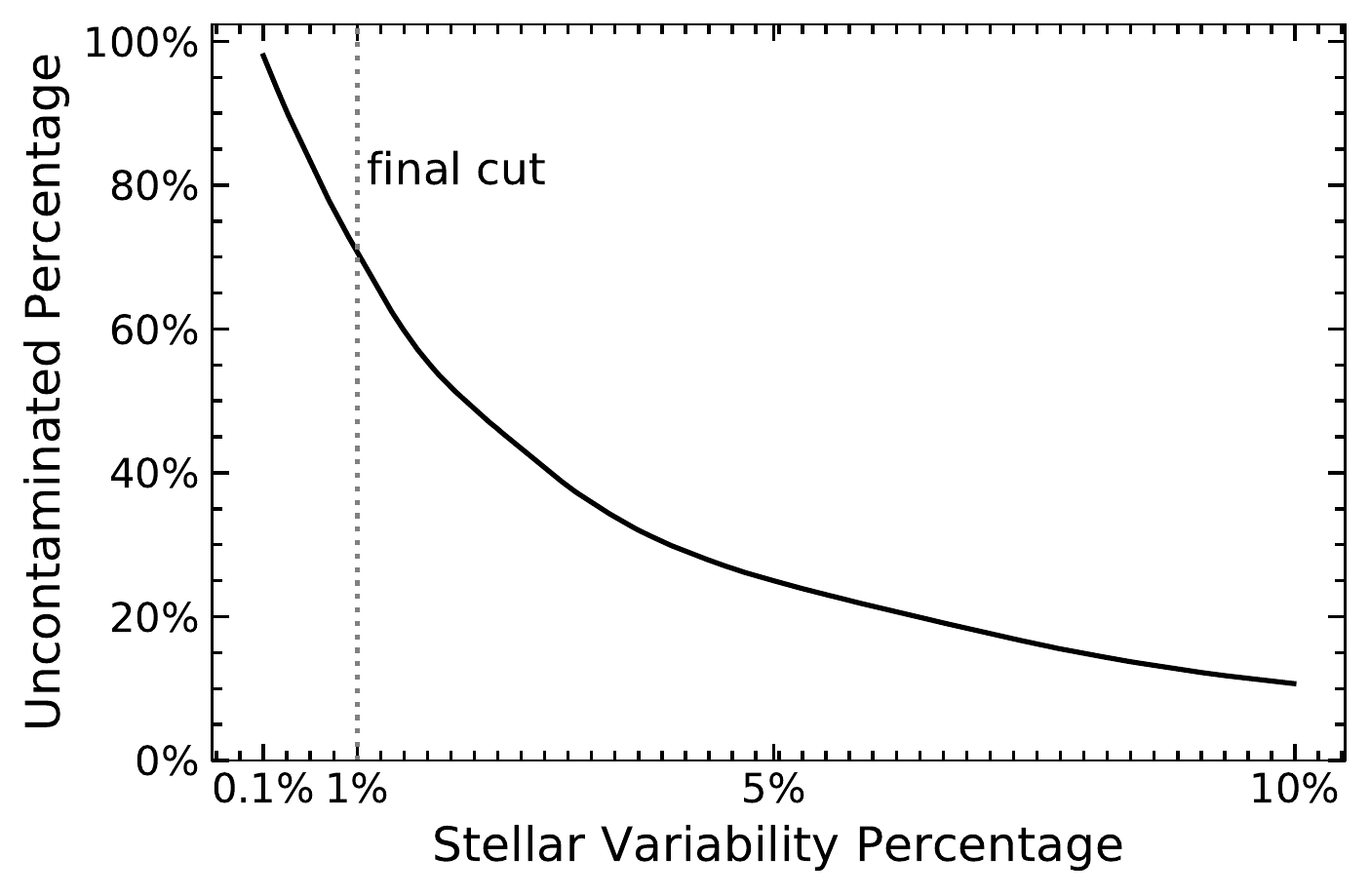}
    \caption{Percentage of sources that we consider ``uncontaminated" by nearby stars cataloged in ATLAS REFCAT2 as a function of the estimated stellar variability percentage attributed to the stars. We only include sources that we consider uncontaminated by the known variable stars in ATLAS-VAR and ASAS-SN.
    We cut at the reasonable value of 1\% variability \citep[e.g.,][]{Huber2011,McQuillen2014,Yu2020}, yielding an ``uncontaminated percentage" of 71\%. Ultimately, though, stellar contamination remains in a significant fraction of the galaxy light curves that are not flagged at this stage.}
    \label{fig:perc}
\end{figure}
\begin{figure*}
    \centering
    \includegraphics[width=2.1\columnwidth]{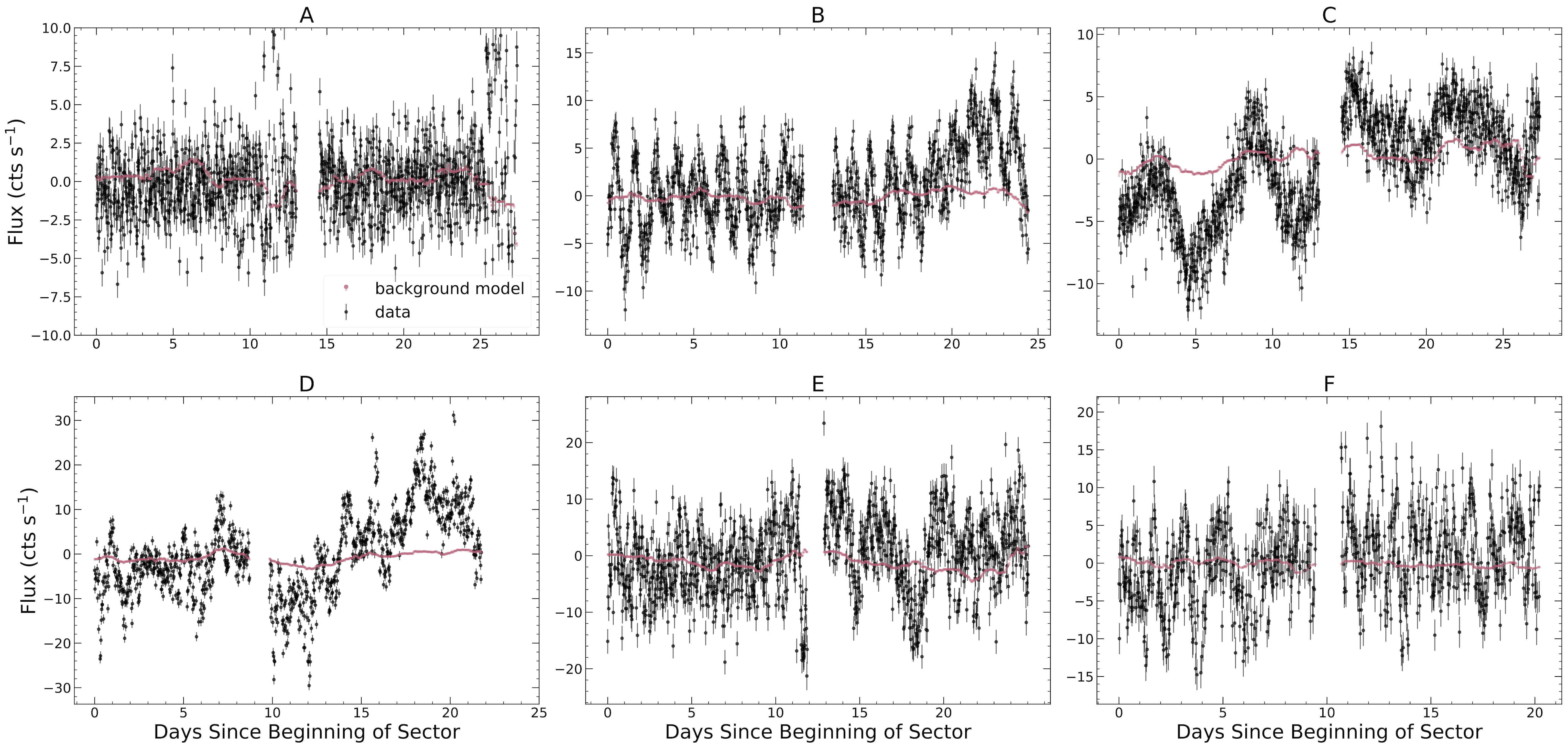}
    \caption{Representative visually inspected light curves are shown in black, while the background model, which has already been subtracted, is shown in pink. 
    A: The light curve is consistent with noise.
    B: The consistent periodicity suggests that this source is likely contaminated by a nearby variable star.
    C: The light curve has too much correlation with the background model to be chosen as a candidate.
    D: This object was cut before visual inspection due to contamination, and is included to illustrate the importance of this step. 
    E: Although NGC 6503 passed visual inspection, we later rejected it because its grid of neighboring light curves showed evidence for systematics that could explain the observed variability.
    F: NGC 0157 is selected as a candidate. The final contamination checks give this source ``low" confidence. The presence of some correlation with the background model is not worrisome, since NGC 0157 is $\sim$4 magnitudes brighter than the source shown in C, while the background model has a much lower relative amplitude.}
    \label{fig:vis}
\end{figure*}
\section{TESS Light Curves}
\label{s.TESS}
We produced TESS light curves following the methods presented in \cite{Fausnaugh2021}. We summarize the steps here. We only included light curves from the primary mission (Sectors 1$-$26).
We employed the difference imaging technique of \cite{Alard1998} and \cite{Alard2000} using the {\tt ISIS} package. 
We used reference images built from 20 full-frame images (FFIs).
These were then scaled in flux and PSF structure and subtracted
from each FFI along with a 2D polynomial model of the sky. The
resulting subtracted images still have many TESS-related systematic
features (e.g., the ``straps'' and various reflection features), which we
remove using median filters to build a model of the residual backgrounds.
The light curves for each source are then extracted.

For sources that passed visual inspection and the RMS check (see Sections \ref{s.rms} and \ref{s.grid}), we also generated light curves according to the techniques outlined in \cite{Vallely2021}. 
This approach is similar to that of \cite{Fausnaugh2021}, but analyzes a local patch around the target source rather than the full FFI using a more involved treatment of various TESS artifacts.
Having light curves from both pipelines allows us to check whether the same AGN-like variability pattern is present in both pipelines. 

\section{Candidate AGN Identification}
\label{s.candidates}
In this section, we describe how we identified the AGN candidates from the sample of 142,061 galaxies.
First, as described in Section \ref{s.sigmag}, we examined the typical light curve magnitude dispersion as a function of magnitude to separate physical variability from noise. Then, we identified nearby stars and calculated whether their estimated variability could account for the apparent variability of the galaxy (Section \ref{s.contamination}). 
Next, we inspected the remaining light curves for AGN-like variability (Section \ref{s.inspection}). The candidates that passed the visual check are listed in Table \ref{t.tab3}. Finally, in Sections \ref{s.rms} and \ref{s.grid}, we describe the further steps we took to check for contamination using inspection of the RMS images and the light curves of nearby pixels. 
\subsection{Sigma vs. Magnitude}
\label{s.sigmag}
To differentiate between noise and physical variability, we examine standard deviation of the light curves in magnitude as a function of the $I$ magnitude (Figure~\ref{fig:prec}). 
Due to crowding and the large TESS pixels, it is difficult to accurately measure a source's mean flux from the TESS data, so we used the cataloged $I$ magnitude because the TESS filter is roughly centered on the $I$-band.

We converted the $I$ band magnitude to TESS counts per second using the TESS filter zero point of 20.44 mag \citep{Vanderspek2018} and then added the counts from the subtracted light curve assuming an average FFI exposure time of 1425.6 seconds (instead of the full 30 minutes) to account for the exposure time lost due to cosmic ray removal.  We then converted back to magnitudes to have a reasonable approximation of a TESS T band (or $I$ band) light curve.

To ensure that the estimated standard deviation was not dominated by TESS systematics, we trimmed each light curve of the commonly problematic parts of a sector. 
In particular, there are often unphysical jumps in count rate at either end of a sector, so we took out the first and last 1.5 days.
There is also a brief interruption in data collection as TESS passes the mid-sector perigee and sends data to the ground. 
We trimmed three days on either side of this downlink because the re-orientation of the spacecraft leads to temperature changes that affect the light curves.
The width of the trimmed time periods are conservative and are used only to compute the standard deviation of the light curves. We keep these regions for our visual inspections.

\subsection{Contamination by Nearby Stars}
\label{s.contamination}
Next, we identified galaxy light curves that may be contaminated by nearby variable stars. We first queried the ATLAS-VAR catalog \citep{ATLAS} and the ASAS-SN Variable Stars Database, which has fewer stars but better coverage of the Southern Hemisphere \citep{Jayasinghe2019,Jayasinghe2021}. We identified stars within 4 arcminutes ($\sim$11 TESS pixels) of the galaxies. The model TESS Pixel Response Function\footnote{https://archive.stsci.edu/missions/tess/models/prf\_fitsfiles/} \citep[PRF,][]{Vanderspek2018} goes to zero at 9 pixels. 
Using a typical PRF and the cataloged variability amplitude, we computed the contribution the variable star would make to the measured flux of the galaxy. We added this ``variable flux" to the estimated Poisson noise in quadrature to obtain an estimate of variability in the galaxy light curve that can be attributed to the star. 
If this value was greater than the light curve standard deviation, we flagged this source as potentially dominated by stellar variability. This step removed 10,805 galaxies.

Ground-based variable star databases usually catalog stars that vary by $\gtrsim1\%$, but TESS is sensitive to lower variability amplitudes. Therefore, we repeated the contamination procedure 
by assigning each nearby ATLAS REFCAT2 \citep{Tonry2018} star a 1\% variability amplitude \citep[e.g.,][]{Huber2011,McQuillen2014,Yu2020}. Of the sources not already flagged using ATLAS-VAR and ASAS-SN, another 29\% are considered contaminated, as shown in Figure \ref{fig:perc}.
A visual inspection of sources \textit{without} this contamination filtering confirms that a significant majority of sources that appear to have AGN-like variability are in fact contaminated.

\begin{deluxetable*}{lDDcDcc}
\tablewidth{0.7\textwidth}
\tabletypesize{\footnotesize}
\tablecaption{AGN Candidates Following Visual Inspection}
\tablehead{
\colhead{Name} &
\multicolumn2c{RA} &
\multicolumn2c{Dec} &
\colhead{\# of Sectors} &
\multicolumn2c{$I$} &
\colhead{log($\mathrm{M_{BH}/M_{\odot}}$)} &
\colhead{Source Outcome}\\
\hline
\colhead{} &
\multicolumn2c{deg} &
\multicolumn2c{deg} &
\colhead{} &
\multicolumn2c{mag} &
\colhead{} &
\colhead{}
}
\decimals
\startdata
NGC1247&48.060 &$-$10.481 &1&11.29 &8.6&grid not AGN-like\\
PGC009923&39.244 &$-$5.349 &1&12.52 &7.9&RMS PRF off-center\\
PGC011248&44.601 &$-$4.295 &1&12.49 &7.1&RMS PRF off-center\\
NGC1232&47.439 &$-$20.579 &1&9.25 &8.3&candidate\\
NGC1365&53.402 &$-$36.141 &1&8.38 &8.7&candidate\\
NGC1395&54.624 &$-$23.027 &1&9.05 &8.6&grid not AGN-like\\
NGC1398&54.717 &$-$26.338 &1&8.39 &8.5&candidate\\
NGC1385&54.368 &$-$24.501 &1&10.13 &7.8&grid not AGN-like\\
NGC1332&51.572 &$-$21.335 &1&9.10 &8.4&candidate\\
PGC014117&58.936 &$-$42.367 &1&10.60 &6.7&RMS PRF off-center\\
\enddata 
\tablecomments{Sources with AGN-like variability identified through visual inspection. Only a section of the table is shown here to demonstrate table formatting; the rest can be found in the ancillary files. We label sources with ``candidate" in the Source Outcome column if they are final AGN candidates. Those that were cut during the RMS check (Section \ref{s.rms}) are labelled ``RMS PRF off-center." We use ``grid not AGN-like" to describe the outcome of sources that pass RMS but not the light curve grid (Section \ref{s.grid}). }
\label{t.tab3} 
\end{deluxetable*}

\subsection{Visual Inspection of Light Curves}
\label{s.inspection}
We visually inspected the 10\% of galaxies that are most variable as well as all $I<13$ mag sources. We looked for aperiodic variability with peaks of varying amplitude.
We simultaneously examined the background model, because the presence of strong common trends, especially when the galaxy is not bright enough for the wings of the galaxy's PRF to affect nearby TESS pixels, implies that a star has contaminated both the galaxy PRF and the background.

We rejected a source during visual inspection if (1) the light curve was consistent with noise with no clear evidence of variability, (2) the light curve appeared clearly consistent with stellar variability because of a periodicity or a strong correlation between the light curve background model, or (3) the light curve variability resembled TESS systematics rather than physical variability (e.g., all the variability was near the sector downlink). We show examples of these various cases in Figure \ref{fig:vis}.
Evolved stars also have quasi-periodic variability with changing peak amplitudes \citep[e.g.,][]{Auge2020}, so we relied on later checks to distinguish between AGNs and irregularly varying stars. The sources that we identified as having AGN-like variability are included in Table \ref{t.tab3}, but further checks for contamination (Sections \ref{s.rms} and \ref{s.grid}) revealed that the large majority of the visually-inspected sources are also contaminated, showing that the contamination filtering in Section \ref{s.contamination} was not sufficiently robust on its own.

\begin{figure*}
    \centering
    \includegraphics[width=1.8\columnwidth]{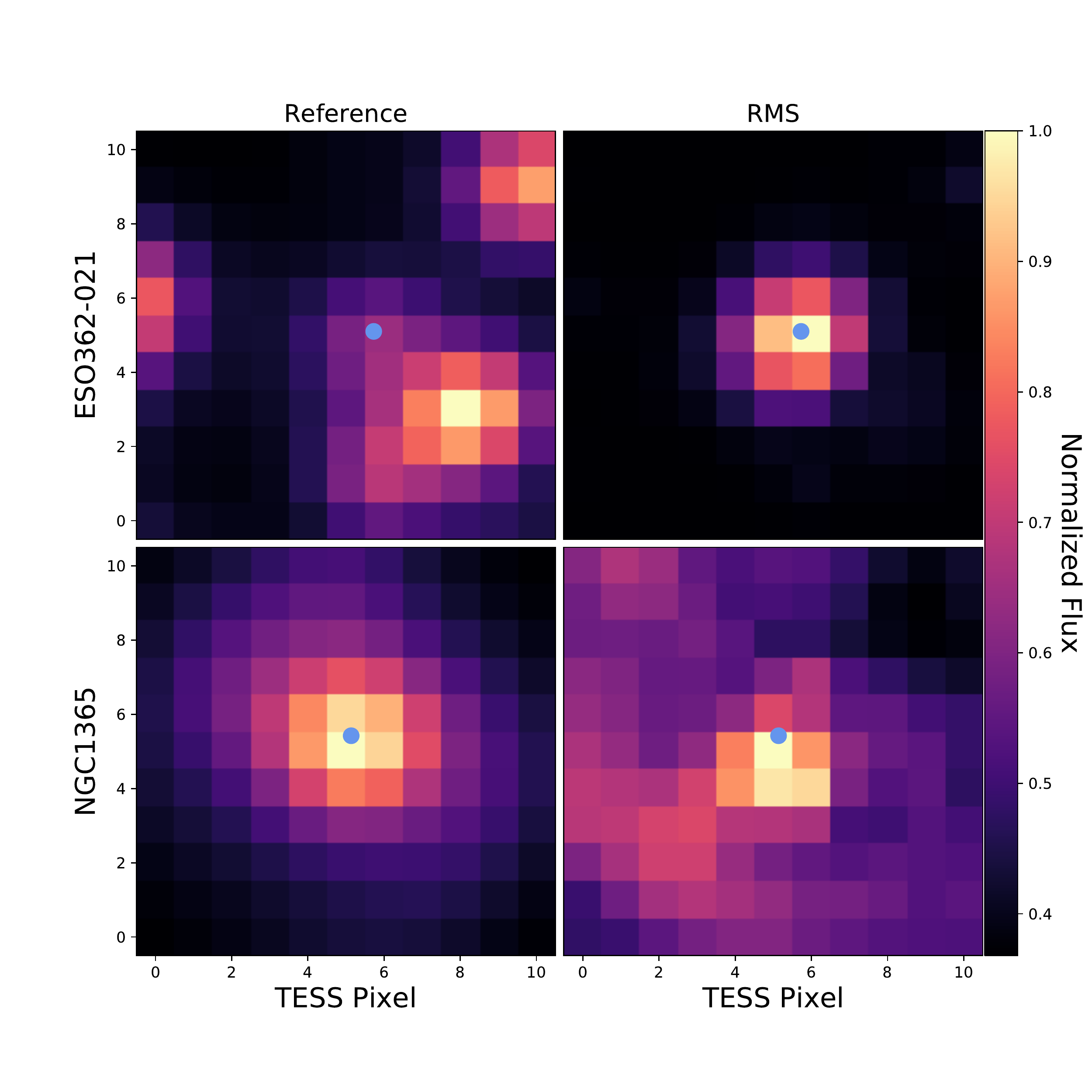}
    \caption{Reference (left) and RMS images (right) of ESO 362-021 (top) and NGC 1365 (bottom), normalized to the brightest pixel. The center of the galaxy is at the blue dot. ESO 362-021 is faint relative to nearby stars, but the RMS image shows that it is the variable source. On the other hand, NGC 1365 is bright but its RMS image has some additional structure and thus calls for further confirmation.}
    \label{fig:rms}
\end{figure*}
\subsection{RMS Images}
\label{s.rms}
For the 258 AGN candidates listed in Table \ref{t.tab3} that passed the visual inspection step (Section \ref{s.inspection}), we next examined the corresponding reference and an RMS image defined by the root mean square of the subtracted images. Peaks in the RMS image correspond to the positions of variable sources and should coincide with the position of the galaxy for a variable AGN. We therefore rejected a source when there was an offset.
Some RMS images are also complicated due to overlapping variable sources or oddly-shaped PRFs. We show examples of the RMS images in Figure \ref{fig:rms}. The top row, which shows the images for ESO 362-021, demonstrates the utility of the RMS images. ESO 362-021 is not the brightest source in the region, but the galaxy is clearly the only variable source in the RMS image.
On the other hand, NGC 1365, in the bottom row of Figure \ref{fig:rms}, has a complicated RMS image. Its reference image shows only the extended bright galaxy. In this case, the reference image is helpful, and the RMS image is acceptable since the variability is centered on the source. 
Of the sources passing visual inspection, only 39 (15\%) have well-centered RMS image sources.

\subsection{Light Curves From Nearby Pixels}
\label{s.grid}
For the objects that passed the RMS check, we performed a two-step final confirmation. We used another TESS light curve pipeline \citep[][]{Vallely2021} to verify the presence of variability. We constructed a $5\times5$ pixel grid of light curves centered on the source pixel of each object.
This grid allowed for a final contamination check.
In particular, we were confident in a signal if the neighboring pixels had the same features as the central light curve but at lower amplitudes, with an expected fall-off consistent with the PRF (e.g., NGC 4395, in Figure 8). One might be concerned that the amplitude at the position of ESO 252-018A (Figure \ref{fig:eso252018a}), for example, is slightly lower than in an adjacent pixel. However, the centering inconsistency is likely an issue with coordinates rather than the source of variability.
In this step, we reject 10 of the 39 sources that passed the RMS check.

\begin{deluxetable*}{lDDccccccccccDccccccccc}
\tabletypesize{\tiny}
\setlength{\tabcolsep}{2.2pt}
\rotate
\tablewidth{0.9\textwidth}
\tablecaption{AGN Candidates}
\tablehead{
\colhead{Name} &
\multicolumn2c{RA} &
\multicolumn2c{Dec} &
\colhead{Conf.} &
\colhead{Est.} &
\colhead{Lit.} &
\colhead{Ref} &
\colhead{Lit.} &
\colhead{Ref} &
\colhead{$\mathrm{log(z)}$} &
\colhead{Dist.} &
\colhead{Ref} &
\colhead{Morph.} &
\multicolumn2c{$I$} &
\colhead{log($\mathrm{L_{\scaleto{X}{2pt}}/erg\:s^{\scaleto{-1}{2pt}}}$)} &
\colhead{W1$-$W2} &
\colhead{W3$-$W4} &
\colhead{Spec.} &
\colhead{Broad} &
\colhead{log([OIII]/H$\beta$)} &
\colhead{log([NII]/H$\alpha$)} &
\colhead{log([SII]/H$\alpha$)} &
\colhead{H$\alpha$ Emission}\\[-0.4cm]&\colhead{}&\multicolumn2c{}&\multicolumn2c{}&
\colhead{log($\mathrm{M_{\scaleto{BH}{2.5pt}}/M_{\scaleto{\odot}{2.5pt}}}$)}&
\colhead{log($\mathrm{M_{\scaleto{\mathrm{BH}}{2.5pt}}/M_{\scaleto{\odot}{2.5pt}}}$)}&
&\colhead{AGN}&\colhead{}&\colhead{}&\colhead{}&\colhead{}&\colhead{}&\colhead{}&\colhead{}&\colhead{}&\colhead{}&\colhead{}&\colhead{Lines}&&&&\colhead{EW}\\
\hline
\colhead{} &
\multicolumn2c{deg} &
\multicolumn2c{deg} &
\colhead{} &
\colhead{} &
\colhead{} &
\colhead{} &
\colhead{} &
\colhead{} &
\colhead{} &
\colhead{Mpc} &
\colhead{} &
\colhead{} &
\multicolumn2c{Vega mag} &
\colhead{} &
\colhead{Vega mag} &
\colhead{Vega mag} &
\colhead{} &
\colhead{} &
\colhead{} &
\colhead{} &
\colhead{} &
\colhead{\AA} }
\decimals
\startdata
IC0356&61.945 &69.812 &high&8.2&7.7&1&\dots&\dots&$-$2.53&$22\pm4$&21&Sb&8.7 &40.6&$0.02\pm0.03$ &$1.31\pm0.05$ &PDS&N&$-0.6\pm0.3$&$0.07\pm0.03$&$-0.49\pm0.08$&$2.4\pm0.1$\\
ESO252-018A&79.957 &$-$45.779 &high&7.7&7.5&2&Sy 2&10&$-$1.46&$154$&z&S0&14.6 &42.6&$0.96\pm0.03$ &$2.27\pm0.03$ &\dots&\dots&\dots&\dots&\dots&\dots\\
ESO362-021&80.742 &$-$36.459 &high&8.3&\dots&\dots&Sy 1&11&$-$1.25&$253$&z&E/S0&13.5 &43.9&$0.98\pm0.03$ &$2.26\pm0.02$ &LDSS3&Y&\dots&\dots&\dots&\dots\\
NGC4395&186.454 &33.547 &high&4.3&5.5&3&Sy 1&13&$-$2.97&$4.3\pm0.4$&22&Sm&12.8 &39.3&$0.79\pm0.03$ &$3.09\pm0.04$ &SDSS&Y&$0.99\pm0.01$&$-0.71\pm0.01$&$-0.45\pm0.01$&$68.2\pm0.6$\\
NGC4449&187.046 &44.094 &high&5.6&3.9&4&\dots&\dots&$-$3.16&$4.3\pm0.4$&23&Sm&9.6 &38.2&$0.16\pm0.03$ &$3.64\pm0.03$ &IIDS&N&$0.5\pm0.0$&$-0.90\pm0.02$&$-0.86\pm0.02$&$27.3\pm0.1$\\
NGC1232&47.439 &$-$20.579 &med.&7.5&6.6&5&\dots&\dots&$-$2.27&$26\pm4$&21&SABc&9.3 &38.7&$-0.05\pm0.03$ &$2.3\pm0.2$ &6dF&N&\dots&$0.42\pm0.15$&$0.15\pm0.22$&$0.5\pm0.1$\\
ESO254-017&91.649 &$-$47.499 &med.&8.7&\dots&\dots&Sy 2&12&$-$1.53&$131$&z&E&11.9 &43.0&$0.21\pm0.03$ &$1.96\pm0.06$ &LDSS3&Y&$0.2\pm0.1$&$0.27\pm0.04$&$-0.29\pm0.09$&$2.8\pm0.2$\\
NGC2654&132.299 &60.221 &med.&7.7&7.7&\dots&\dots&\dots&$-$2.35&$32\pm4$&24&SBab&11.3 &\dots&$-0.03\pm0.03$ &$1.39\pm0.08$ &SDSS&N&$0.47\pm0.08$&$0.26\pm0.04$&$-0.011\pm0.051$&$0.71\pm0.05$\\
NGC2683&133.173 &33.422 &med.&7.5&7.4&4&LINER&14&$-$2.86&$9.4\pm0.3$&25&Sb&8.6 &39.0&$0.09\pm0.03$ &$1.55\pm0.04$ &PDS&N&$0.28\pm0.05$&$0.21\pm0.02$&$0.022\pm0.026$&$2.6\pm0.1$\\
UGC04767&136.439 &36.355 &med.&8.1&\dots&\dots&\dots&\dots&$-$1.62&$106$&z&S0&12.5 &\dots&$-0.08\pm0.03$ &$1.8\pm0.1$ &SDSS&N&$-0.008\pm0.254$&$-0.036\pm0.102$&$-1.0\pm0.6$&$0.57\pm0.09$\\
NGC2759&137.155 &37.622 &med.&8.1&\dots&\dots&\dots&\dots&$-$1.64&$103\pm19$&26&E-S0&12.4 &\dots&$-0.06\pm0.03$ &$1.2\pm0.5$ &SDSS&N&$0.027\pm0.090$&$0.032\pm0.072$&$-1.0\pm0.6$&$0.63\pm0.07$\\
NGC2782&138.521 &40.114 &med.&7.8&8.0&4&Sy 1&15&$-$2.07&$37\pm14$&27&SABa&10.9 &39.9&$0.48\pm0.03$ &$2.72\pm0.02$ &Bok&N&$0.083\pm0.010$&$-0.24\pm0.01$&$-0.42\pm0.01$&$21.0\pm0.1$\\
ESO536-014&354.760 &$-$25.670 &low&7.9&\dots&\dots&\dots&\dots&$-$1.50&$142\pm31$&28&Sb&12.6 &\dots&$-0.06\pm0.03$ &$2.1\pm0.1$ &6dF&N&$0.27\pm0.29$&$0.007\pm0.105$&$-0.68\pm0.47$&$1.8\pm0.3$\\
NGC55&3.723 &$-$39.197 &low&5.3&\dots&\dots&\dots&\dots&$-$3.36&$1.86\pm0.08$&29&Sm&8.7 &37.5&$0.08\pm0.04$ &$3.57\pm0.04$ &\dots&\dots&\dots&\dots&\dots&\dots\\
NGC0157&8.695 &$-$8.396 &low&7.2&7.7&5&LINER&16&$-$2.26&$12\pm2$&24&SABb&9.5 &\dots&$0.03\pm0.03$ &$2.59\pm0.03$ &\dots&\dots&\dots&\dots&\dots&\dots\\
PGC089900&22.199 &$-$54.357 &low&8.3&\dots&\dots&Sy 1&17&$-$1.03&$426$&z&S&14.4 &44.6&$1.06\pm0.03$ &$2.87\pm0.03$ &\dots&\dots&\dots&\dots&\dots&\dots\\
NGC0720&28.252 &$-$13.739 &low&8.6&7.7&6&\dots&\dots&$-$2.24&$26\pm3$&24&E&9.3 &39.2&$-0.03\pm0.03$&$1.12\pm0.08$&\dots&\dots&\dots&\dots&\dots&\dots\\
NGC959&38.100 &35.495 &low&5.4&5.5&4&\dots&\dots&$-$2.70&$11\pm2$&24&Sm&11.6 &\dots&$0.01\pm0.04$ &$2.91\pm0.06$ &Bok&N&$0.22\pm0.02$&$-0.56\pm0.02$&$-0.35\pm0.01$&$6.91\pm0.07$\\
NGC0991&38.886 &$-$7.154 &low&6.0&\dots&\dots&\dots&\dots&$-$2.29&$19\pm4$&27&SABc&12.1 &39.2&$-0.04\pm0.04$ &$2.5\pm0.2$ &SDSS&N&$0.3\pm0.1$&$0.041\pm0.044$&$0.04\pm0.04$&$1.08\pm0.07$\\
UGC02509&45.850 &4.492 &low&7.3&\dots&\dots&\dots&\dots&$-$1.70&$79\pm15$&24&Sbc&13.3 &\dots&$0.12\pm0.03$ &$1.99\pm0.05$ &SNIFS&N&\dots&$-0.20\pm0.07$&$-0.58\pm0.15$&$1.7\pm0.1$\\
NGC1332&51.572 &$-$21.335 &low&8.0&8.8&7&\dots&\dots&$-$2.27&$25\pm3$&24&E-S0&9.1 &39.3&$-0.03\pm0.03$ &$1.22\pm0.05$ &\dots&\dots&\dots&\dots&\dots&\dots\\
NGC1365&53.402 &$-$36.141 &low&8.0&6.0&5&CLAGN&18&$-$2.26&$18.3\pm0.5$&30&Sb&8.4 &40.2&$0.76\pm0.03$ &$3.41\pm0.02$ &6dF&Y&$0.23\pm0.01$&$-0.28\pm0.01$&$-1.04\pm0.02$&$34.5\pm0.2$\\
NGC1398&54.717 &$-$26.338 &low&8.3&8.0&8&\dots&\dots&$-$2.33&$29\pm6$&24&SBab&8.4 &40.6&$0.01\pm0.03$&$1.20\pm0.08$&\dots&\dots&\dots&\dots&\dots&\dots\\
NGC2146&94.657 &78.357 &low&7.7&7.3&4&AGN&19&$-$2.53&$17\pm3$&27&SBab&9.6 &39.4&$0.60\pm0.03$ &$2.97\pm0.02$ &Bok&N&$-0.23\pm0.02$&$-0.23\pm0.01$&$-0.55\pm0.01$&$15.70\pm0.05$\\
NGC2976&146.815 &67.916 &low&5.9&5.1&4&\dots&\dots&$-$5.00&$3.6\pm0.1$&31&Sc&9.2 &38.6&$0.04\pm0.03$ &$2.88\pm0.04$ &SDSS&N&$0.13\pm0.01$&$-0.52\pm0.01$&$-0.69\pm0.01$&$73.1\pm0.6$\\
NGC3183&155.454 &74.177 &low&7.3&\dots&\dots&\dots&\dots&$-$1.99&$37\pm7$&24&SBbc&11.0 &\dots&$0.27\pm0.03$ &$2.17\pm0.03$ &\dots&\dots&\dots&\dots&\dots&\dots\\
NGC4244&184.374 &37.807 &low&5.7&5.7&9&\dots&\dots&$-$3.09&$4.4\pm0.2$&32&Scd&10.9 &37.0&$-0.04\pm0.03$ &$2.22\pm0.12$ &SDSS&N&$-0.18\pm0.13$&$-0.47\pm0.04$&$-0.23\pm0.04$&$3.0\pm0.1$\\
NGC7552&349.045 &$-$42.584 &low&7.7&7.3&5&LINER&20&$-$2.27&$20\pm4$&27&Sab&9.3 &40.8&$0.79\pm0.03$ &$3.12\pm0.02$ &6dF&N&$-0.65\pm0.04$&$-0.12\pm0.01$&$-0.66\pm0.03$&$16.6\pm0.1$\\
ESO605-016&354.279 &$-$20.463 &low&7.8&\dots&\dots&\dots&\dots&$-$1.58&$115$&z&Sc&11.7 &\dots&$-0.11\pm0.03$ &$2.3\pm0.3$ &SNIFS&Y&\dots&$-0.16\pm0.03$&$-0.73\pm0.07$&$1.5\pm0.1$\\
\enddata 
\tablecomments{Final Candidates. The references for literature $\mathrm{M_{BH}}$ are as follows: (1) is \cite{Richings2011},
(2) is \cite{Lewis2006},
(3) is \cite{denBrok2015},
(4) is \cite{Williams2021},
(5) is \cite{Davis2014},
(6) is \cite{MutluPakdil2016},
(7) is \cite{Barth2016},
(8) is \cite{Saglia2016}, and
(9) is \cite{deLorenzi2013}. 
The literature mass estimates generally employ higher-accuracy methods than our own. However, the \cite{Williams2021} value of $\mathrm{log(M_{BH}/M_{\odot})} = 3.91\pm0.81$ for NGC 4449 has significant uncertainty, and uses an extrapolation of the $\mathrm{M-\sigma}$ of \citet{Tremaine2002}. In this case, our $\mathrm{log(M_{BH}/M_{\odot})}$ estimate of 5.6 may be more representative.
The references for sources selected as AGNs in the literature are as follows: 
(10) is \cite{Glass1981},
(11) is \cite{Westerlund1966,Danziger1979},
(12) is \cite{Pietsch1998},
(13) is \cite{Filippenko1989},
(14) is \cite{Spinoglio1989},
(15) is \cite{Balzano1983},
(16) is \cite{Pulatova2015},
(17) is \cite{Maza1994},
(18) is \cite{Giorgio2003,Risaliti2000},
(19) is \cite{Tzanavaris2007,Ruschel-Dutra2017}, and
(20) is \cite{Dudik2007}. The distance references are as follows: (21) is \cite{Tullypeculiar},
(22) is \cite{NGC4395dist},
(23) is \cite{NGC4449dist},
(24) is \cite{cosmicflows3},
(25) is \cite{NGC2683dist},
(26) is \cite{NGC2759dist},
(27) is \cite{Tully1988},
(28) is \cite{cosmicflows4},
(29) is \cite{NGC55dist},
(30) is \cite{NGC1365dist},
(31) is \cite{cosmicflows2}, and 
(32) is \cite{NGC4244dist}. Distance references marked as ``z" are converted from the given redshifts using the cosmological parameters adopted by HyperLEDA: $H_{0} = 70 \: \mathrm{km \: s^{-1} \: Mpc^{-1}}$, 
$\Omega_{M} = 0.27$, and
$\Omega_{\Lambda} = 0.73$. For these conversions, we used \cite{Wright2006}.
We took available galaxy morphologies from HyperLEDA. Morphologies for ESO 362-021 and PGC 089900 were estimated using DSS images.
For candidates from the GLADE subsample, we identified morphologies in \cite{deVaucouleurs1991}.
When possible, we used the $I$ magnitudes cataloged in HyperLEDA. However, for NGC 4395 and NGC 55, we estimated $I$ using the $J$ mag measurements cataloged in GLADE (Section \ref{s.sample}), and the NGC 959 value is from \cite{Heraudeau1996}.
Estimated X-ray luminosity (see Section \ref{s.xray}) is in the 0.5--10 keV band.
The ``Spec." column provides the name of the telescope and/or spectrograph used for the emission-line diagnostic diagrams (Figure \ref{fig:bptwhan}). 
We use LDSS3 to designate the Low Dispersion Survey Spectrograph 3 on the Magellan Clay telescope, IIDS for the Intensified Image Dissector Scanner on the Kitt Peak 2.1-m telescope, PDS to refer to 
the Palomar Double Spectrograph, and Bok for the Boller \& Chivens Spectrograph on the 2.3-m Bok Telescope.}
\label{t.fcand} 
\end{deluxetable*}

\begin{deluxetable}{lDDc}
\tablewidth{0.7\textwidth}
\tabletypesize{\footnotesize}
\tablecaption{Source Outcomes for Entire Sample}
\tablehead{
\colhead{Name} &
\multicolumn2c{RA} &
\multicolumn2c{Dec} &
\colhead{Reason For Cut} \\
\hline
\colhead{} &
\multicolumn2c{deg} &
\multicolumn2c{deg} &
\colhead{}}
\decimals
\startdata
PGC092785&346.139 &$-$32.677 &contaminated (REFCAT2)\\
PGC095282&0.408 &$-$27.566 &contaminated (ATLAS-VAR)\\
PGC100817&340.217 &$-$28.308 &contaminated (ASAS-SN)\\
UGC03014&64.974 &2.093 &low variability\\
ESO117-016&59.200 &$-$60.427 &low variability\\
PGC014487&61.802 &$-$17.205 &low variability\\
PGC096217&355.918 &-12.478 &not AGN-like\\
ESO603-011&341.467 &$-$20.842 &RMS PRF off-center\\
NGC6667&277.666 &67.987 &grid not AGN-like\\
NGC0991&38.886 &$-$7.154 &candidate\\
\enddata 
\tablecomments{Names (for HyperLEDA), coordinates, and the source outcome for the whole sample (HyperLEDA and GLADE). First, sources removed due to expected stellar contamination (Section \ref{s.contamination}) are designated as ``contaminated" with a specification of the database of origin for the star of concern (REFCAT2, ATLAS-VAR, or ASAS-SN Variable Stars Database). A ``Reason For Cut" of ``low variability" describes candidates that passed the contamination cut but have $I>$13 mag and are not in the top 10\% variable in their magnitude bin (Figure \ref{fig:prec}, Section \ref{s.inspection}). Next, ``not AGN-like" indicates that a candidate passed the variability cut but did not have AGN-like variability (Section \ref{s.inspection}). A final stage of ``RMS PRF off-center" designates that a candidate has potential AGN-like variability but does not have the variable PRF in the RMS image centered on the galactic nucleus (Section \ref{s.rms}). Sources with ``grid not AGN-like" have a centered PRF in the RMS image, but the grid of neighboring light curves from the independent reduction (Section \ref{s.grid}) suggested that the variability does not originate from the galaxy. Lastly, the final candidates are described by ``candidate."}
\label{t.final} 
\end{deluxetable}

\begin{figure*}
    \centering
    \includegraphics[width=1.9\columnwidth]{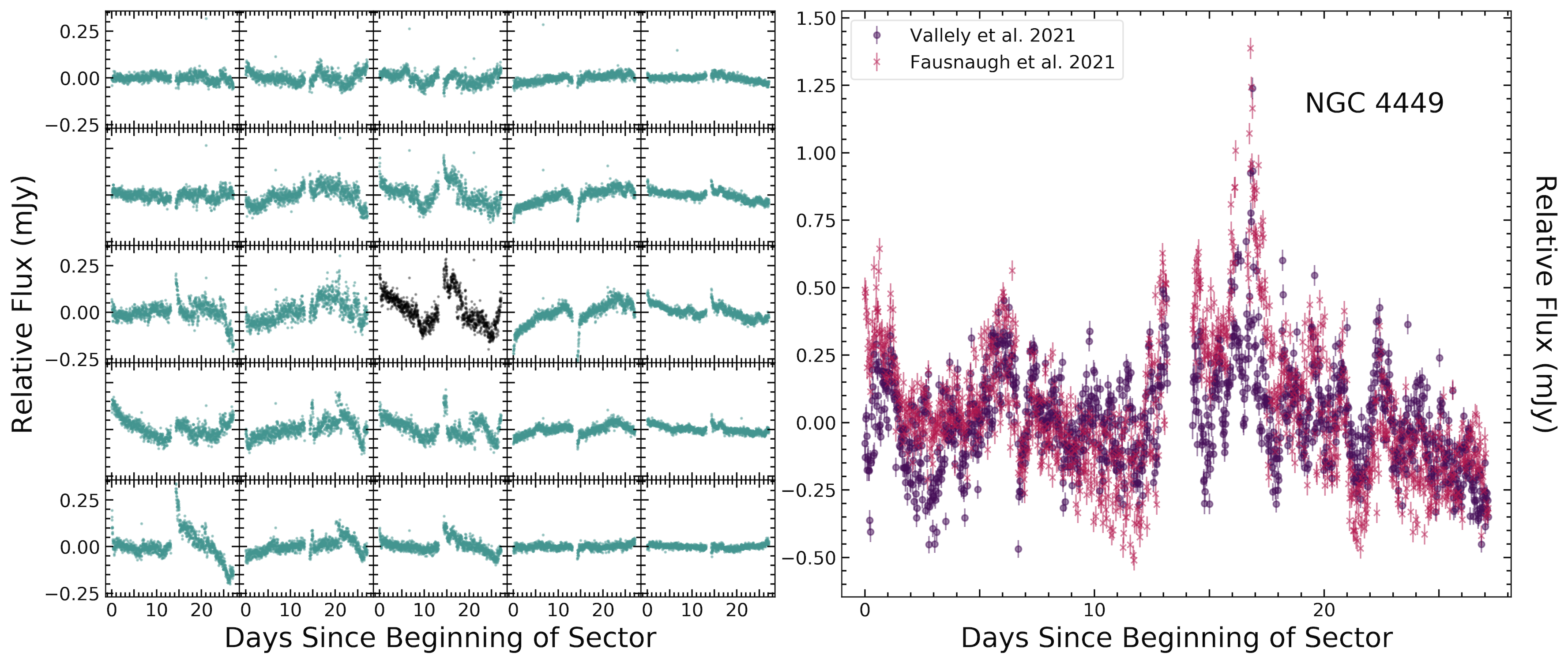}
    \caption{Final confirmation step for the low-mass, high-confidence source NGC 4449. The grid of Sector 22 light curves (left) are for the source's central and neighboring pixels using the \citet[][]{Vallely2021} pipeline. The corresponding PSF photometry version is shown with pink x's (right), with the original reduction used in the visual inspection shown in purple.}
    \label{fig:ngc4449}
\end{figure*}
\begin{figure*}
    \centering
    \includegraphics[width=1.9\columnwidth]{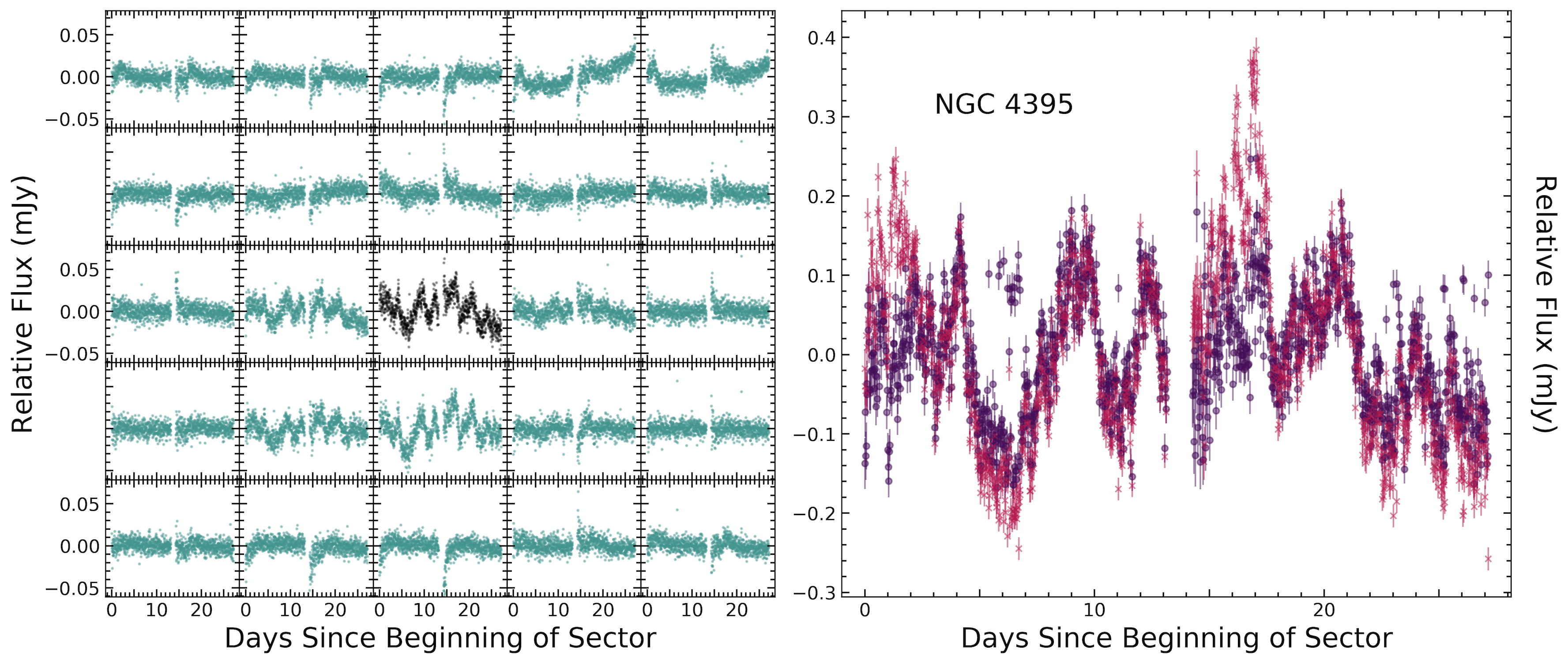}
    \caption{Same as Figure \ref{fig:ngc4449}, but for high-confidence candidate NGC 4395.}
    \label{fig:ngc4395}
\end{figure*}
\begin{figure*}
    \centering
    \includegraphics[width=1.9\columnwidth]{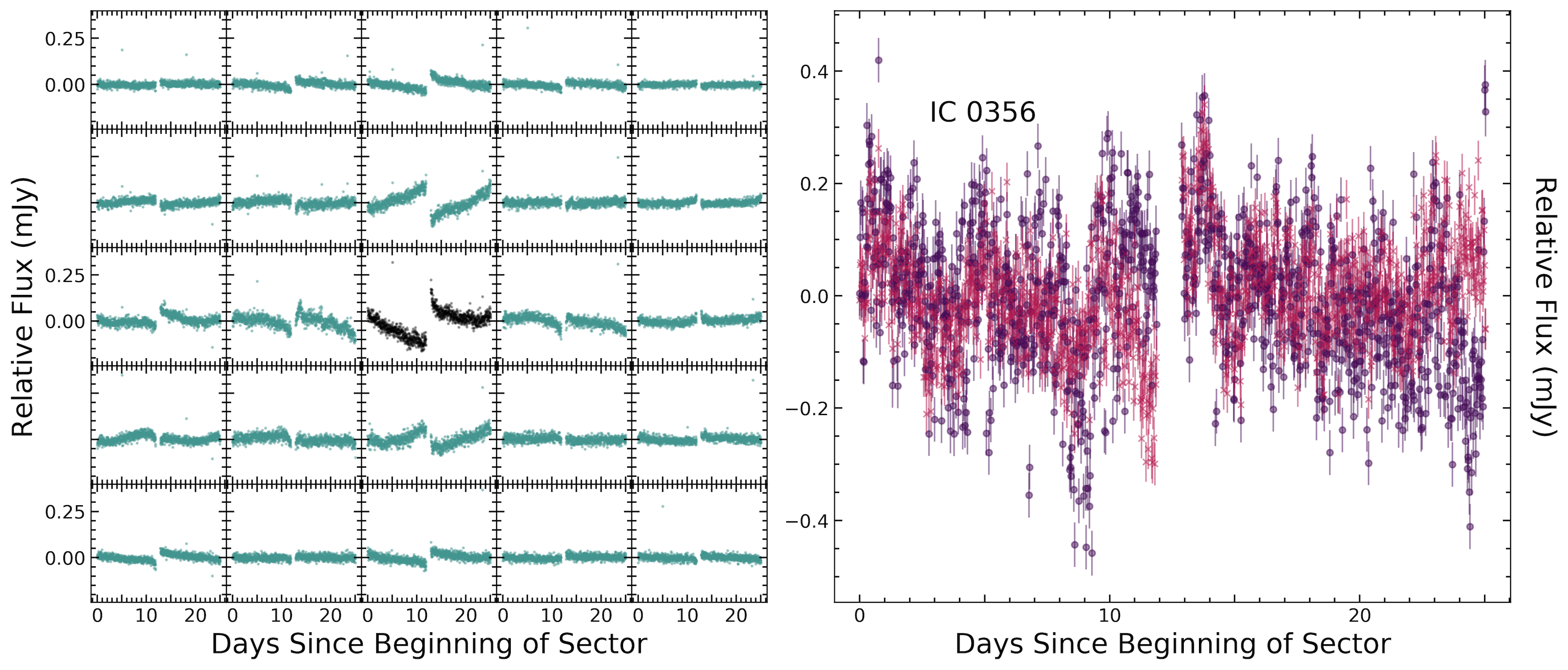}
    \caption{Same as Figure \ref{fig:ngc4449}, but for high-confidence candidate IC 0356 in Sector 19. The discontinuity at the time of a data downlink in the pixel grid light curves is fairly typical.}
    \label{fig:ic0356}
\end{figure*}
\begin{figure*}
    \centering
    \includegraphics[width=1.9\columnwidth]{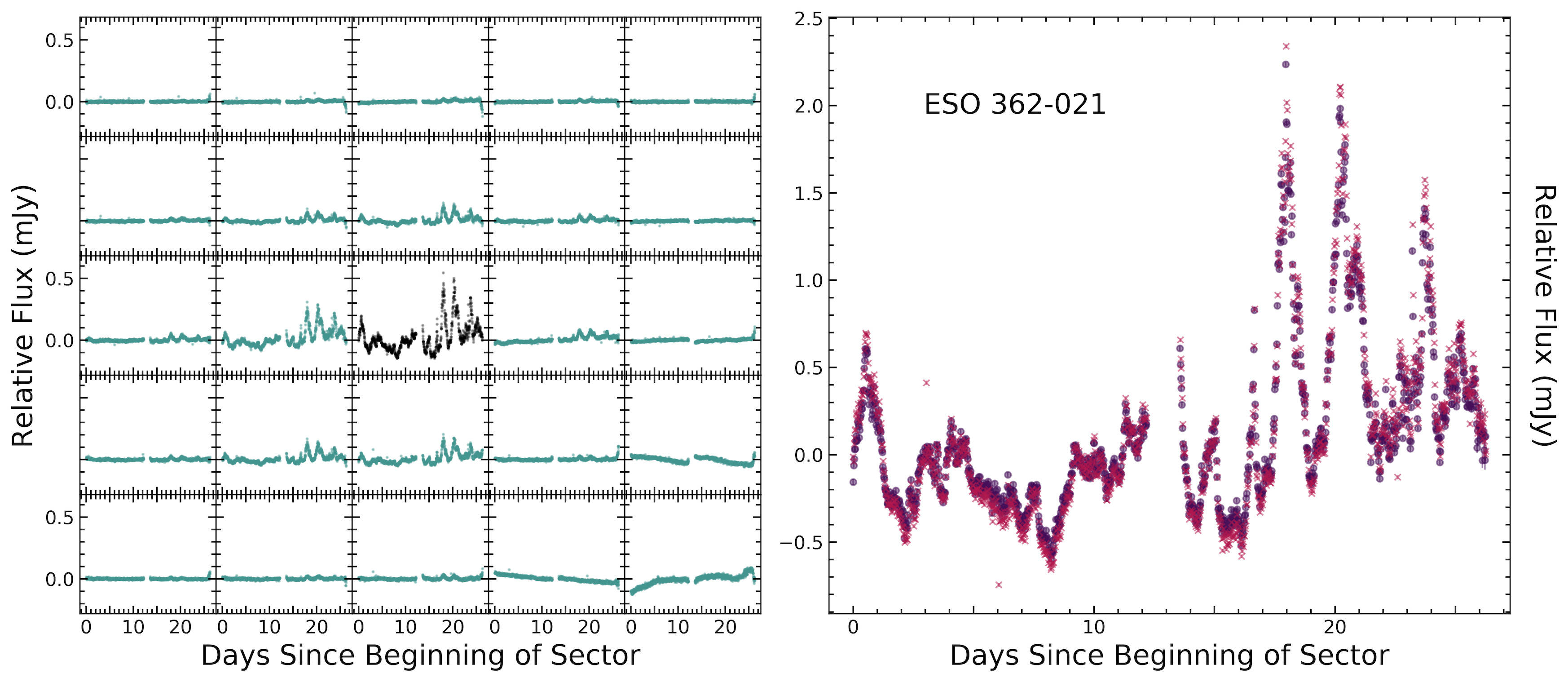}
    \caption{Same as Figure \ref{fig:ngc4449}, but for high-confidence candidate ESO 362-021 in Sector 5.}
    \label{fig:eso362021}
\end{figure*}
\begin{figure*}
    \centering
    \includegraphics[width=1.9\columnwidth]{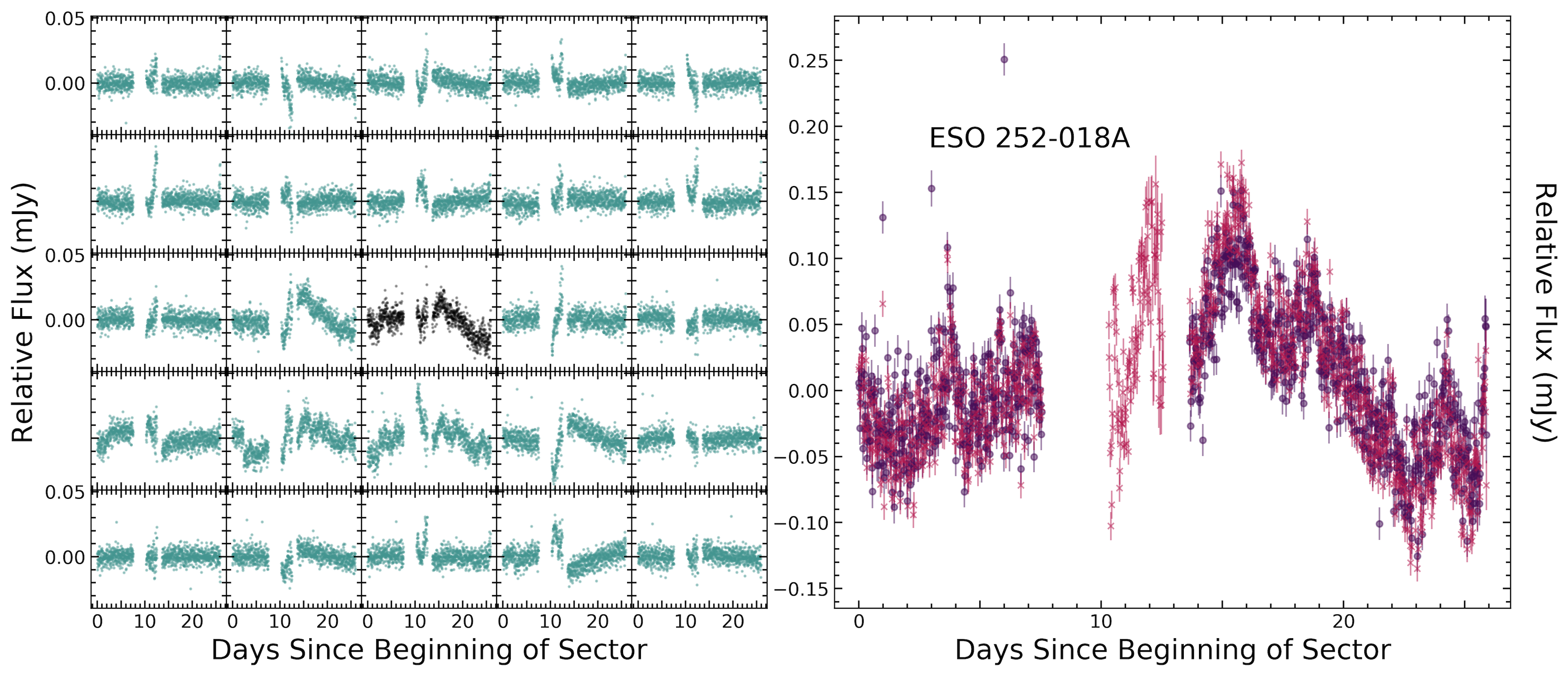}
    \caption{Same as Figure \ref{fig:ngc4449}, but for high-confidence candidate ESO 252-018A in Sector 4. The stricter rejection procedure in the reduction methods outlined in \citet{Vallely2021} yielded a different temporal range for this light curve and a few others.}
    \label{fig:eso252018a}
\end{figure*}
\begin{figure*}
    \centering
    \includegraphics[width=1.9\columnwidth]{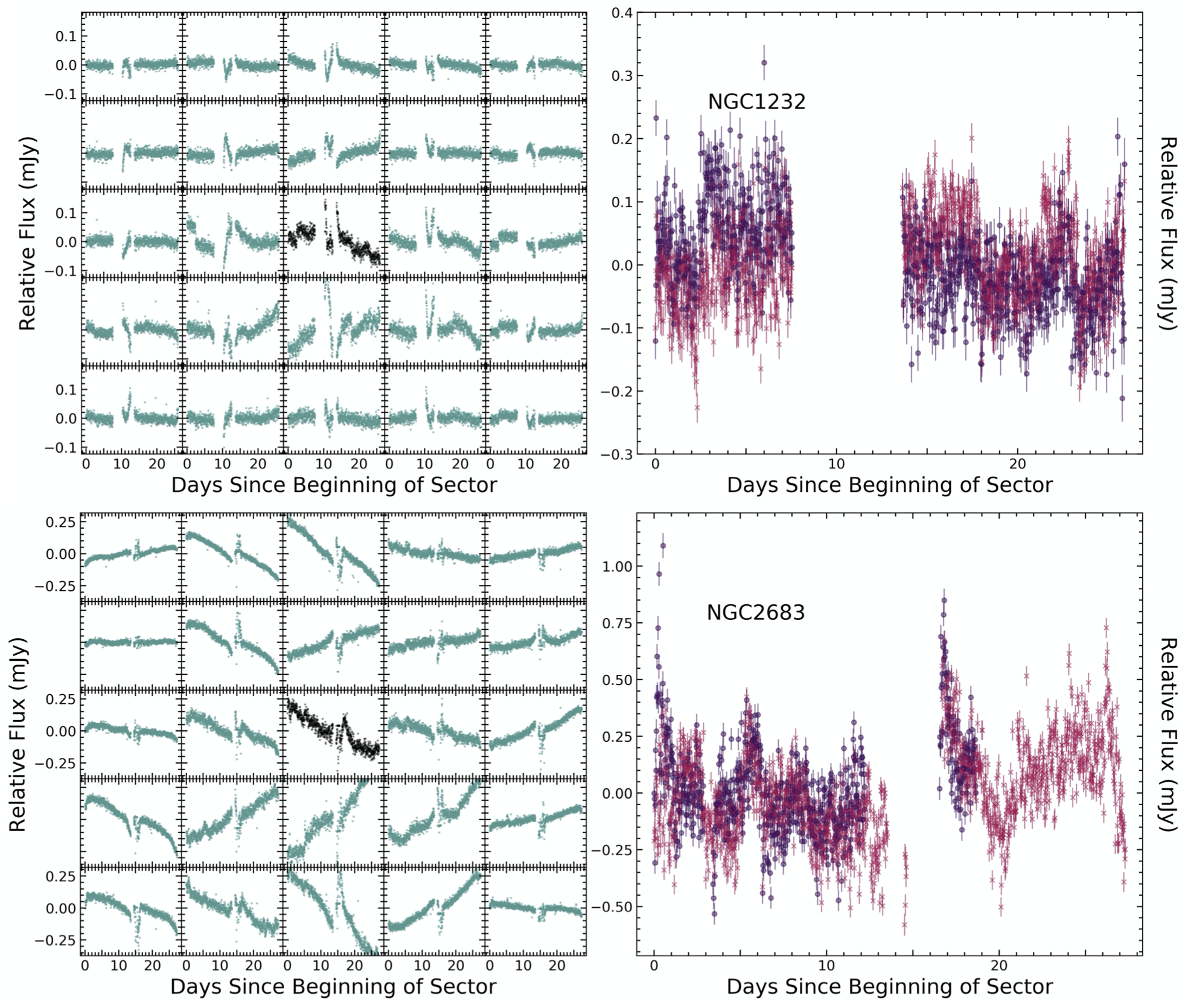}
    \caption{Same as Figure \ref{fig:ngc4449}, but for representative medium-confidence candidates. For each source, there is decent agreement between the two reductions.
    \textit{Top}: Sector 4 light curves of NGC 1232. The short-term variations are strongest in the center of the grid, and there is an expected decrease in amplitude in the neighboring light curves.
    \textit{Bottom}: Sector 21 light curves of NGC 2683. The light curves from the neighboring pixels have significant long-term trends, which are likely systematics, but the short-term signal from the central pixel is present in an expected way in the neighboring light curves.}
    \label{fig:medium}
\end{figure*}
\begin{figure*}
    \centering
    \includegraphics[width=1.9\columnwidth]{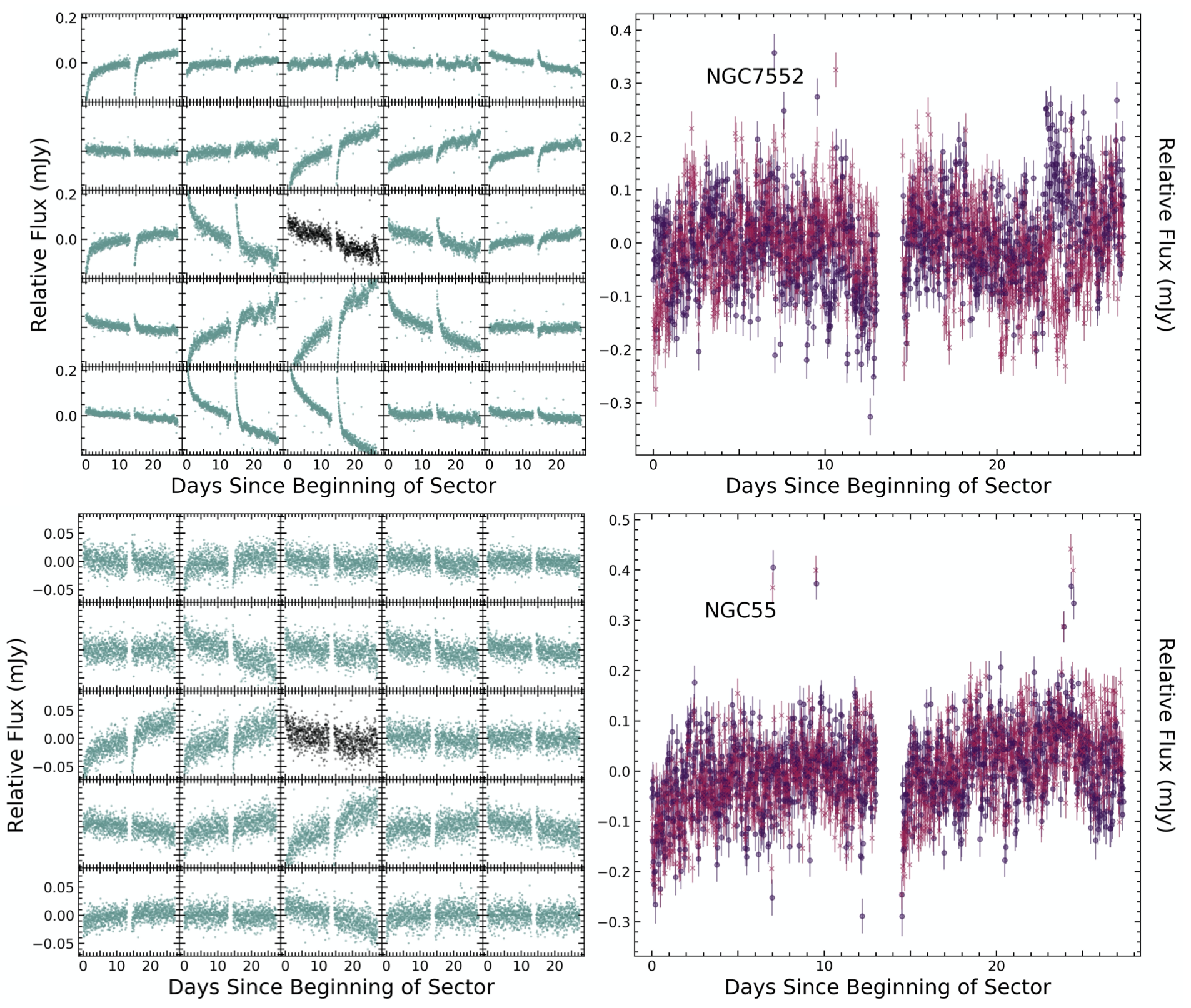}
    \caption{Same as Figure \ref{fig:ngc4449}, but for representative low-confidence candidates. \textit{Top}: The pixels surrounding the Sector 2 light curve of NGC 7552 show similar behavior to those of NGC 2683 (Figure \ref{fig:medium}). NGC 7552 is lower-confidence, because of the decreased detectability of similar, but lower-amplitude, variability in the neighboring light curves. \textit{Bottom}: The Sector 2 light curves of NGC 55 show a weak signal, leading to its designation as low-confidence.}
    \label{fig:low}
\end{figure*}

\subsection{Final Candidates}
\label{s.finalcands}
The final 29 AGN candidates are presented in Table \ref{t.fcand} and Table \ref{t.final} summarizes the outcomes for each source in the original HyperLEDA and GLADE samples. 
Based on the RMS images and light curve grids, we assign each final candidate one of the three confidence levels.
High-confidence sources show no instrumental issues or stellar contamination. They also have similar, lower-amplitude trends in the nearby pixels.
In Figures \ref{fig:ngc4449}$-$\ref{fig:eso252018a} we show the light curve grids for our five high-confidence sources.
Medium-confidence sources show some evidence of systematic problems or nearby stars, but likely not enough to explain the observed variability. We show light curve grids for two medium-confidence sources in Figure \ref{fig:medium}. Low-confidence sources may be AGNs, but there are enough indications of systematic errors or contamination to potentially explain the observed variability. In Figure \ref{fig:low} we show representative examples of low-confidence sources. In total, there are 5 high-confidence sources, 7 medium-confidence sources, and 17 low-confidence sources.

Because the original BH mass estimates were rough, we attempted to re-estimate $\mathrm{M_{BH}}$ for each of the final AGN candidates. First, we searched the literature for black hole mass estimates and found 18 estimates which are included in Table \ref{t.fcand}. Most sources only had one literature estimate, but for those with more, we chose estimates in the order: (1) spatially-resolved kinematics, (2) virial mass, and (3) stellar velocity dispersion.
Second, we found morphology classifications for the candidates using the NASA Extragalactic Database (NED).\footnote{The NASA/IPAC Extragalactic Database (NED) is operated by the Jet Propulsion Laboratory, California Institute of Technology, under contract with the National Aeronautics and Space Administration.} 
For the HyperLEDA and GLADE subsamples, we originally took the $\mathrm{M_{BH}}$ upper limits using the \cite{Graham2007} relation. By incorporating the galaxy morphologies, we scaled down the absolute $K$-band measurements to bulge magnitudes using the median scaling factors found for each type in \cite{Graham2008}. 
Third, using NED and the Extragalactic Distance Database \citep[EDD,][]{Tully2009}, we found redshift-independent distances for 23 of the candidates. We gave preference to the distances in the order: (1) the Cepheid period-luminosity relation, (2) tip of the red-giant branch (TRGB), (3) supernovae Ia, (4) the Tully-Fisher relation, and (5) surface brightness fluctuations.
For the six candidates without redshift-independent distance, we 
used their spectroscopic redshifts to estimate luminosity distances.
For the remainder of this work, we use the literature $\mathrm{M_{BH}}$ where available, and otherwise employ our re-estimated $K$-band values.

\section{Classifying Candidates Using Other AGN Diagnostics}
\label{s.diagnostics}
In this section, we compare our TESS variability selection to other methods of AGN selection.
In Section \ref{s.veron}, we describe our investigation of the source outcomes of all the galaxies in the HyperLEDA sample that are labeled as AGNs.
We then used archival and new data to identify which objects can be selected as AGNs using other methods, including line diagnostics (Section \ref{s.emission}), color-color diagrams (Section \ref{s.wise}), and X-ray luminosity selection techniques (Section \ref{s.xray}). Table \ref{t.fcand} includes the results of this search.

If a source is not found to be an AGN using methods other than variability, that does not rule out the possibility. 
In \cite{Yuk2022}, emission-line diagnostic diagrams classify $\sim$30\% of variability-selected AGN candidates as star-forming. For low-mass AGNs, spectroscopic selection misses 75\% of variability-selected candidates in \cite{Baldassare2020} and 81\% in \cite{Ward2021}.

\subsection{AGNs in HyperLEDA Not Selected With Variability}
\label{s.veron}
HyperLEDA includes nuclear activity classifications primarily taken from \cite{Veron2006}.
Among the sources that passed the stellar contamination check, 231 are cataloged as AGNs and 57 are selected as variables using the brightness or standard deviation cut. 
Following visual inspection, six of these AGNs remain. The six objects are in our final candidate list. 
Thus, although variability selection using TESS identifies AGNs missed by other methods, it selects only a minority of known AGNs.

\subsection{Emission-Line Diagnostic Diagrams}
\label{s.emission}
\begin{figure*}
    \centering
    \includegraphics[width=2\columnwidth]{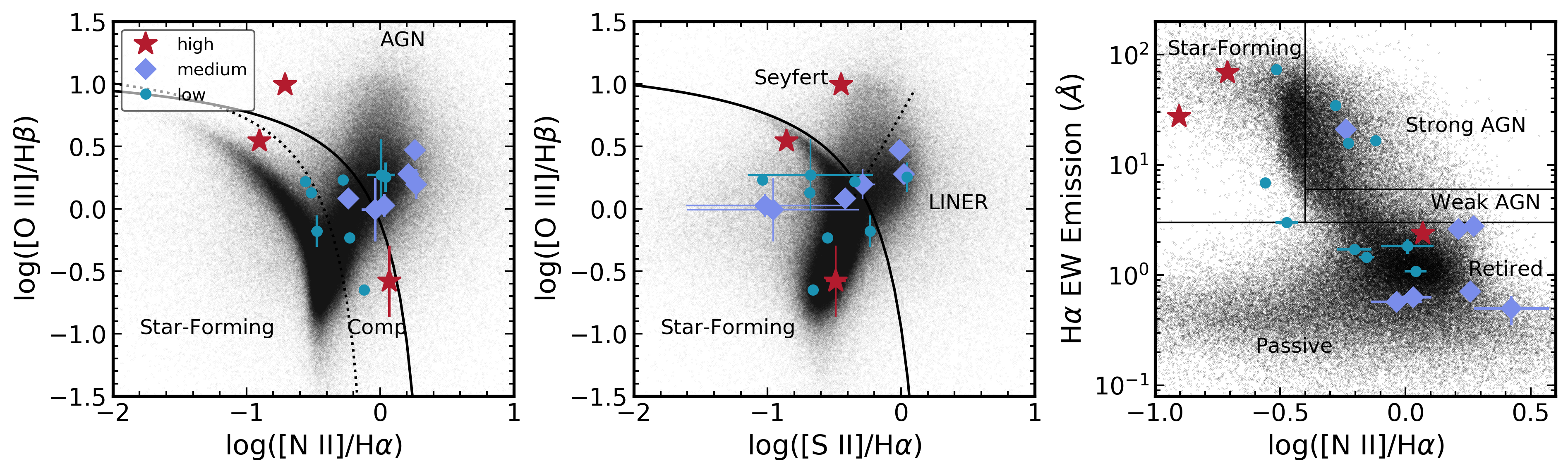}
    \caption{Emission-line diagnostic diagrams for the AGN candidates, demonstrating that variability selection allows for identification of AGNs missed by other methods.  
    The high-confidence candidates are shown by red stars, the medium-confidence by purple diamonds, and the low-confidence by blue circles. The black points are galaxies with spectra in SDSS Data Release 8 \citep[][]{SDSSDR8}. \textit{Left}: $\mathrm{log([O III]/H\beta)\:vs.\:log([N II]/H\alpha)}$, which is one of the BPT/VO87 diagrams \citep{Baldwin1981,Veilleux1987}.
    The solid curve was theoretically
    determined in \cite{Kewley2001} to separate star-forming regions and AGN. The dotted curve follows the empirical separation from \cite{Kauffmann2003}.
    \textit{Center}: $\mathrm{log([O III]/H\beta)\:vs.\:log([S II]/H\alpha)}$ \citep{Baldwin1981,Veilleux1987}. The solid curve originates from \cite{Kewley2001}, while the dotted line theoretically differentiating between Seyferts and LINERs comes from \cite{Kewley2006}.
    \textit{Right}: $\mathrm{H\alpha\:EW\:vs.\:log([N II]/H\alpha)}$  diagram \citep[WHAN,][]{Fernandes2011}. 
    Of the 17 candidates present in all three diagrams, 76\% are selected as AGNs by at least one of these metrics, but none are selected by all three. Table \ref{t.bpt} provides more specific percentages by diagram.}
    \label{fig:bptwhan}
\end{figure*}
The BPT/VO87 diagrams \citep{Baldwin1981,Veilleux1987,Kewley2006} are common methods for distinguishing between AGNs and galaxies with star formation. We examined $\mathrm{log([O III]/H\beta)}$ vs.$\:\mathrm{log([N II]/H\alpha)}$ and $\mathrm{log([O III]/H\beta)}$ vs.$\:\mathrm{log([S II]/H\alpha)}$. 
The forbidden line distinction is also used in the WHAN diagram \citep[$\mathrm{EW_{H\alpha}}$ vs. $\mathrm{[NII]/H\alpha}$,][]{Fernandes2010,Fernandes2011}. This classification method requires fewer lines and selects a higher proportion of galaxies as AGNs.

We used NED to obtain optical spectra from the Sloan Digital Sky Survey \citep[SDSS][]{SDSS}, 
6dF \citep{6dF}, the Palomar Double Spectrograph \citep[][]{Oke1982}, the Boller \& Chivens Spectrograph on the 2.3-m Bok Telescope,\footnote{http://james.as.arizona.edu/$\sim$psmith/90inch/bcman/html/bcman.html}
and the Intensified Image Dissector Scanner on the Kitt Peak 2.1-m telescope \citep[][]{IIDS}.
We also searched for spectra using the Gemini Observatory, NOIRLab, European Southern Observatory, Keck, and Las Cumbres Observatory \citep[][]{Brown2013} archives. 
We also obtained two spectra using the Low Dispersion Survey Spectrograph 3\footnote{http://www.lco.cl/technical-documentation/index-2/} (LDSS3) on the Magellan Clay telescope at Las Campanas Observatory, and
two using the Supernova Integrated Field Spectrograph (SNIFS;  \citealp{lantz04}) on the University of Hawai'i 88-inch telescope as part of the Spectral Classification of Astronomical Transients (SCAT; \citealp{tucker18}) survey. 

In total, we found or acquired spectra for 21 of the 29 candidates. Using \texttt{pPXF} \citep{Cappellari2017}, we were able to fit the lines of 18 of these spectra (excluding $H\beta$ for three of the 18) and used the SDSS line fits \citep[][]{Brinchmann2004} for two additional objects.
Table \ref{t.fcand} includes the origin of each spectrum. In Figure \ref{fig:bptwhan}, the AGN candidates are highlighted in the BPT and WHAN diagrams according to the associated confidences.

Of the 17 variability-selected AGN candidates for which we could use all three emission-line diagnostic diagrams, 13 (76\%) are selected as AGNs by at least one of the diagrams, but none are selected by all three.
In addition, the LDSS3 spectrum of ESO 362-021 and the SNIFS spectrum of ESO 605-016 had unconstrained line fits but clearly exhibit broad lines.
Table \ref{t.bpt} summarizes these spectroscopic characterizations.
These results demonstrate the strength of our methodology in two ways. First, the selection percentage is significantly higher than the overall AGN fraction (5-20\% depending on the method), further confirming the physicality of the detections. Second, the emission-line diagnostic methods do not select some of our candidates, thus corroborating that variability selection is crucial for population studies of AGN.
\begin{table}
\caption{AGN Candidate Selection in Emission-Line Diagnostic Diagrams (Figure \ref{fig:bptwhan})} 
\begin{tabularx}{\columnwidth}{llrr}
\hline\hline
\multicolumn{2}{c}{$\mathrm{[OIII]/H\beta}$ vs. $\mathrm{[NII]/H\alpha}$} \\
\hline
AGN 
&  1 (5.9\%) & 
LINER 
&  7 (41.1\%) \\ 
Composite 
& 5 (29.4\%) \\ 
Star-Forming 
&  4 (23.5\%) &\\
\hline
\hline
\multicolumn{2}{c}{$\mathrm{[OIII]/H\beta}$ vs. $\mathrm{[SII]/H\alpha}$} \\ 
\hline
Seyfert 
& 1 (5.9\%) &
LINER 
& 4 (23.5\%) \\ 
Star-Forming 
&  12 (70.6\%) &\\
\hline
\hline
\multicolumn{2}{c}{$\mathrm{EW_{H\alpha}}$ vs. $\mathrm{[NII]/H\alpha}$}\\ 
\hline
Strong AGN 
&  4 (20.0\%) & 
Weak AGN 
&  0 (0.0\%)  \\ 
Passive or Retired 
&  12 (60.0\%) & 
Star-Forming 
&  4 (20.0\%) \\
\hline
\end{tabularx}
\tablecomments{There are 17 objects in the $\mathrm{[OIII]/H\beta}$ vs. $\mathrm{[NII]/H\alpha}$ and $\mathrm{[OIII]/H\beta}$ vs. $\mathrm{[SII]/H\alpha}$ diagrams, and 20 objects in the $\mathrm{EW_{H\alpha}}$ vs. $\mathrm{[NII]/H\alpha}$ (WHAN) plot.}
\label{t.bpt}
\end{table}
\subsection{X-ray Detections}
\label{s.xray}
We queried the individual catalogs included in the results of a HEASARC Master X-ray Catalog\footnote{https://heasarc.gsfc.nasa.gov/W3Browse/all/xray.html} search: ASCAGIS \citep{ascagis1,ascagis2}, BMWCHANCAT \citep{bmwchancat}, BMWHRICAT \citep{bmwhricat}, CHAMPPSC \citep{champpsc}, CXOXASSIST \citep{cxoxassist}, EINGALCAT \citep{eingalcat}, RASS2RXS \citep{rass2rxs}, RASS6DFGS \citep{rass6dfgs}, RASSBSCPGC \citep{rassbscpgc}, TARTARUS \citep{tartarus}, WGACAT \citep{wgacat}, XMMSLEWCLN \citep{xmmslewcln}, XMMSSC \citep{xmmssc}, and XMMSTACK \citep{xmmstack}. 
We converted the count rates to the flux in the 0.5--10 keV band using the Portable, Interactive Multi-Mission Simulator \citep[PIMMS,][]{Mukai1993}. We used a power-law model with photon index 1.75 \citep[e.g.,][]{Ricci2017} and included the Galactic absorption.\footnote{https://heasarc.gsfc.nasa.gov/cgi-bin/Tools/w3nh/w3nh.pl} 
Where possible, we retrieved inputs for PIMMS from the catalogs or their documentation. For XMMSTACK we assumed that the medium filter was used throughout. Likewise, for the ROSAT PSPC catalogs, we used the open filter. 

\begin{figure}
    \centering
    \includegraphics[width=\columnwidth]{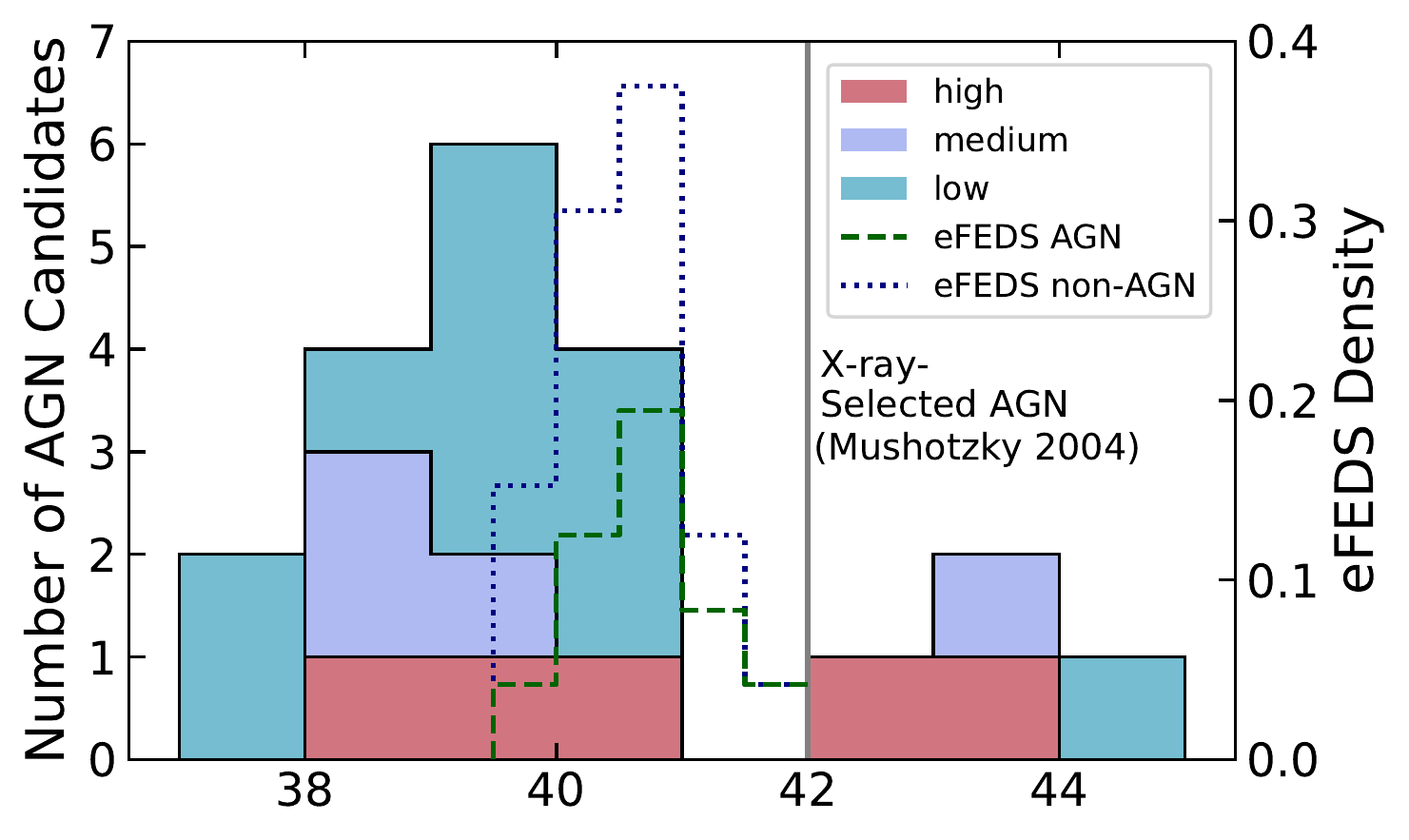} 
    \caption{The filled stacked histograms show the median X-ray luminosity for the 20 detected AGN candidates by confidence level.
    All 5 of our high-confidence candidates have X-ray detections.
    The X-ray luminosity cutoff from \cite{Mushotzky2004} marks the value at which the source of the emission must be an AGN rather than star formation. This cutoff yields four X-ray selected AGN, while the \citet[][]{Latimer2021} criterion for low-mass galaxies selects three additional sources.
    The unfilled stacked histogram shows the X-ray luminosity distribution for HyperLEDA galaxies within 200 Mpc detected in the eROSITA Final Equatorial Depth Survey (eFEDS), with the AGN and non-AGN populations separated with emission line diagnostic diagrams \citep{Vulic2022}.}
    \label{fig:xhist}
\end{figure}

Figure \ref{fig:xhist} shows the distribution of median X-ray luminosities for the 20 candidates with catalog matches 
as well as the distributions for AGN (Seyfert, LINER, and composite galaxies) and non-AGN detected in the eROSITA Final Equatorial Depth Survey \citep[eFEDS,][]{Brunner2022} field \citep[][]{Vulic2022}. 
Of the 1181 HyperLEDA galaxies within 200 Mpc in the eFEDS footprint, \citet{Vulic2022} identified 94 matches, 76 of which could be classified using optical spectra. We estimated the 0.5--10 keV X-ray luminosity using the 0.5--2 keV luminosities from \citet{Vulic2022} and the conversion factor from \citet[][]{Comparat2022}.

Using the common luminosity cutoff of $\mathrm{L_{X}>10^{42}}$ erg $\mathrm{s^{-1}}$ \citep[e.g.,][]{Mushotzky2004} to exclude sources where the X-ray emission could plausibly originate from an X-ray binary population, we identify
four objects (ESO 252-018A, ESO 254-017, ESO 362-021, and PGC 089900) as X-ray AGNs.
However, the maximum X-ray binary contribution is lower for lower-mass galaxies so a limit of $10^{39-40} \mathrm{\: erg \: s^{-1}}$ can be used  \citep[e.g.,][]{Latimer2021}, adding three low-mass galaxies (NGC 0991, NGC 1365, and NGC 4395).
Therefore, X-ray luminosity and optical variability select overlapping but different portions of AGN parameter space, as is the case with other selection methods.

\subsection{WISE Color-Color Cuts}
\label{s.wise}
\begin{figure}
    \centering
    \includegraphics[width=\columnwidth]{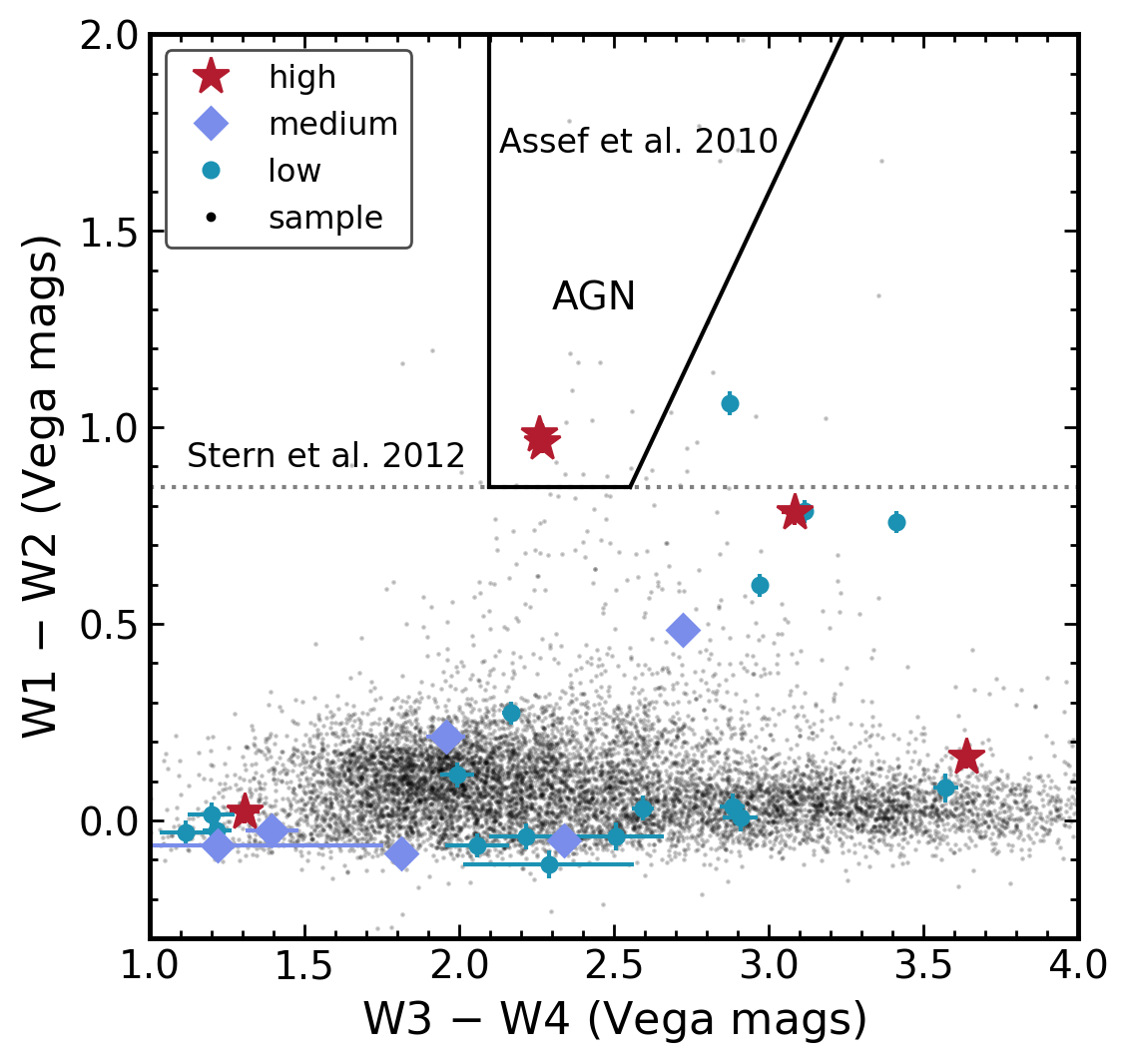}
    \caption{W1$-$W2 vs. W3$-$W4 for candidate AGNs with WISE color measurements. The high-confidence candidates are shown by red stars, the medium-confidence by purple diamonds, and the low-confidence by blue circles. 10,000 random HyperLEDA sources that are at least as close as farthest AGN candidate are plotted in black.
    Two candidates are selected as an AGN using the wedge from \cite{Assef2010}, but a third galaxy is selected using the criterion from \cite{Stern2012}, which only uses W1$-$W2. All three are also X-ray selected (Section \ref{s.xray}). 
    }
    \label{fig:wise}
\end{figure}

For galaxies in which an AGN dominates the mid-IR emission the system will be redder in mid-IR color than stars or galaxies with weak AGNs \citep{Assef2010,Stern2012,Assef2013}.
Figure \ref{fig:wise} shows the the Wide-Field Infrared Survey Explorer \citep[WISE,][]{Wright2010} W1$-$W2 and W3$-$W4 colors of the candidates. Only ESO 362-021 and ESO 252-018A are selected as AGNs using the wedge described in \citet[][]{Assef2010,Assef2013}. These two objects also have maximum X-ray luminosities above the selection limit described in Section \ref{s.xray}.
However, other studies \citep[e.g.,][]{Stern2012} select AGNs using just a W1$-$W2 cutoff of 0.8 mag which adds the X-ray selected candidate PGC 089900.
These results are consistent with the low fraction of MIR-selected AGNs in other studies comparing selection methods \citep[e.g.,][]{Laurent2000}. \\
\begin{figure*}
    \centering
    \includegraphics[width=2\columnwidth]{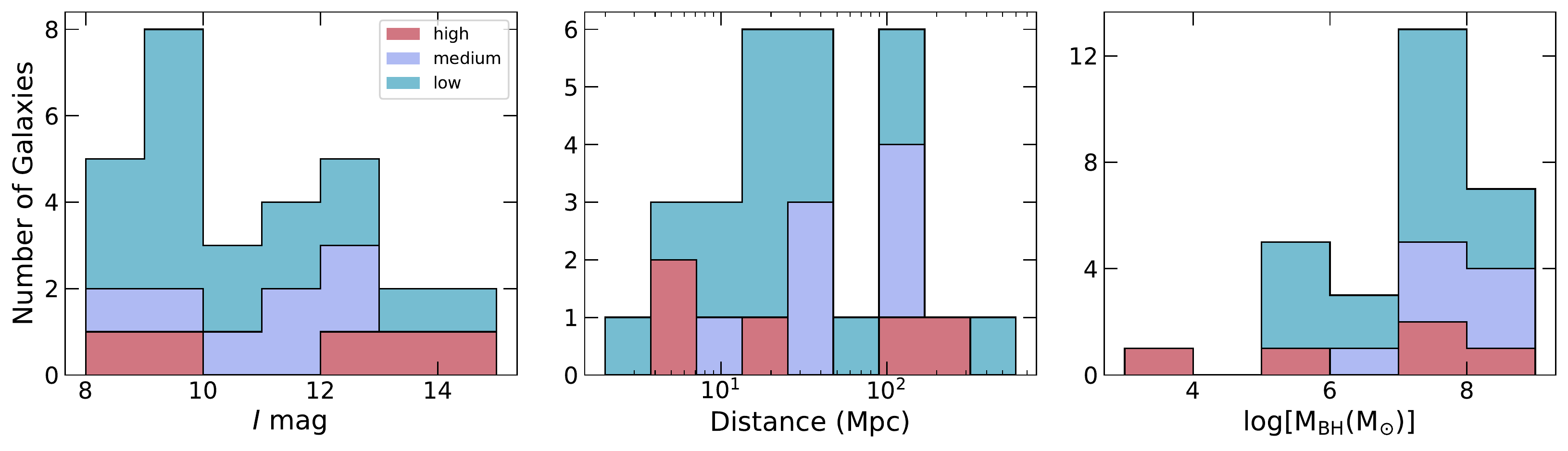}
    \caption{Histograms of apparent $I$-band magnitude, distance, and estimated black hole mass for the AGN candidates. Each histogram is stacked by candidate confidence. These histograms can be compared to those of the original sample (Figure \ref{fig:hists}). The literature mass estimate for NGC 4449 of $3.9\pm0.89$ $\mathrm{M_{\odot}}$ \cite{Williams2021} is shown. The uncertainty is significant and the estimate requires an extrapolation of the $\mathrm{M-\sigma}$ relation from \citet{Tremaine2002}. We note that our corresponding estimate is $\mathrm{log(M_{BH}/M_{\odot})}=5.6$ using the \cite{Graham2008} and \cite{Graham2007} relations to convert the galaxy absolute $K$-band magnitude to the spheroid absolute $K$-band magnitude, followed by the $\mathrm{log(M_{BH})}$.}
    \label{fig:chists}
\end{figure*}
\section{Conclusions}
\label{s.conclusion}
We searched for strongly variable AGNs using TESS light curves of 142,061 galaxies. We selected our sample from two sources. First, we used HyperLEDA sources in nine TESS sectors to identify AGNs without restrictions on $\mathrm{M_{BH}}$. Second, we considered an all-sky sample of 4366 low-mass galaxies from GLADE. Within these samples, we identify 29 optically-varying AGN candidates, of which 18 are newly-identified. Although the GLADE sample comprised only 3\% of the total sample, 17\% of the final AGN candidates originated from GLADE.

The primary goal of our study was the identification of variable low-mass AGNs with TESS. 
Figure \ref{fig:chists} shows the distribution of estimated black hole mass for the AGN candidates. 
Of the 29 candidates, 8 have estimated $\mathrm{M_{BH}}$ $\mathrm{\lesssim10^{6}M_{\odot}}$ (see Table \ref{t.fcand}).
Of the low-mass galaxies, two are known AGNs.
Three of the seven low-mass galaxies with available spectra are selected as AGNs in at least one of the emission-line diagnostic diagrams (Figure \ref{fig:bptwhan}).
Overall, 76\% of the candidates with available spectra are selected as AGNs in at least one line ratio diagram. In this percentage, we include composite and LINER galaxies, as they likely include the contribution of some AGN light \citep[e.g.,][]{Kewley2006}.

We independently selected NGC 4395 as the strongest low-mass candidate, corroborating the \cite{Burke2020} identification of its TESS variability.
NGC 4395 is a particularly close dwarf AGN (d = 4.3 Mpc). We estimated the distance limit at which the variability of NGC 4395 would be detectable by TESS. Using Figure \ref{fig:prec}, we found that NGC 4395 would be dominated by noise if it were one magnitude fainter. This corresponds to sources $\sim$1.4 times more distant and thus the volume probed by TESS for comparable levels of variability to NGC 4395 is $\sim 6$ Mpc. A few of our low-mass AGN candidates are more distant than 10 Mpc, but they are all more luminous than NGC 4395.

Most of our candidates are higher-mass AGNs cataloged in HyperLEDA. Single TESS sectors can be long enough for the identification of variability in AGNs with estimated black hole masses up to $\mathrm{\sim}$$\mathrm{10^{9} M_{\odot}}$, which would be expected to correspond to a $\mathrm{\tau_{DRW}}$ of $\gtrsim$500 days according to the correlation in \cite{Kelly2009}. 

For GLADE objects, we generated light curves for any relevant primary mission sector, whereas the HyperLEDA light curves originate from just nine of the 26 TESS sectors. A natural next step would be an investigation of the HyperLEDA galaxies in the remaining sectors, including the extended missions, to corroborate candidate AGNs and identify additional candidates that were not included in our sample or were caught in a lower state. 

Retrospectively, we would have made use of the RMS images earlier in the paper because of the extensive stellar variable contamination created by TESS' low resolution.
Most seemingly AGN-like variability in TESS is in fact stellar in origin. However, 74\% of visually-inspected candidates that passed the RMS check (Section \ref{s.rms}) remained candidates following the final confirmation step of examining neighboring light curves (Section \ref{s.grid}). The RMS images are useful as an indication of both variability amplitude and position.
In particular, one should use centroiding to identify sources with variability consistent with the galaxy nucleus. Furthermore, with a growing sample of example AGN candidate light curves, it is more straightforward to turn the techniques of visual inspection (Section \ref{s.inspection}) into an algorithmic approach.

Assuming a constant rate of HyperLEDA AGN candidates in each TESS primary mission sector, we roughly expect to identify 45 additional AGN candidates with similarly strong variability amplitudes, of which $\sim$26 will be new and $\sim$8 will be in low-mass galaxies. This sample will enable further examination of the $\mathrm{\tau_{DRW}-M_{BH}}$ relation, as well as of the underlying integrity of the DRW model. 

\section*{Acknowledgements}
HT acknowledges support from Research Experience for Undergraduate program at the Institute for Astronomy, University of Hawaii-Manoa funded through NSF grant \#2050710.
HT would also like to thank the Institute for Astronomy for their hospitality during the course of this project.

JTH was supported by NASA grant 80NSSC21K0136. BJS and CSK are supported by NSF grant AST-1907570/AST-1908952. BJS is also supported by NSF grants AST-1920392 and AST-1911074. CSK is also supported by NSF grant AST-181440.
XD is supported by NASA grant 80NSSC22K0488.

This research has made use of the NASA/IPAC Extragalactic Database (NED),
which is operated by the Jet Propulsion Laboratory, California Institute of Technology,
under contract with the National Aeronautics and Space Administration.

We acknowledge the usage of the HyperLeda database (http://leda.univ-lyon1.fr).

This research has made use of data obtained from XMMSL2, the Second XMM-Newton Slew Survey Catalogue, produced by members of the XMM SOC, the EPIC consortium, and using work carried out in the context of the EXTraS project (``Exploring the X-ray Transient and variable Sky", funded from the EU's Seventh Framework Programme under grant agreement no. 607452).

\bibliography{newdraft.bib}{}
\bibliographystyle{aasjournal}

\end{document}